\documentclass[fleqn,usenatbib]{mnras}

\usepackage[T1]{fontenc}
\usepackage{ae,aecompl}

\DeclareRobustCommand{\VAN}[3]{#2}
\let\VANthebibliography\thebibliography
\def\thebibliography{\DeclareRobustCommand{\VAN}[3]{##3}\VANthebibliography}

\usepackage{graphicx}	
\usepackage{amsmath}	

\usepackage{epstopdf}
\usepackage{adjustbox}
\usepackage{textcomp, gensymb}
\usepackage{subcaption} 
\usepackage{tabu}
\usepackage{float}
\usepackage{rotfloat}
\usepackage{rotating}
\usepackage{multirow}
\usepackage{multicol}
\usepackage{pdflscape} 
\usepackage{breqn} 
\usepackage[flushleft]{threeparttable}
\usepackage{orcidlink}
\usepackage{array}
\usepackage{amsmath}
\newcommand{\plus}{\scalebox{1.1}{+}}
\newcommand{\ie}{i.e.,}
\newcommand{\eg}{e.g.,}

\title[Short title, max. 45 characters]{Hard X-ray view of Compton-thick AGN in Circinus Galaxy}

\title{Multi-epoch hard X-ray view of Compton-thick AGN Circinus Galaxy}

\author[Kayal et al.]{
Abhijit Kayal $^{\orcidlink{0000-0001-9851-8243}}$,$^{1,2}$\thanks{E-mail: abhijitk@prl.res.in}
Veeresh Singh $^{\orcidlink{0000-0002-6040-4993}}$,$^{1}$
Claudio Ricci $^{\orcidlink{0000-0001-5231-2645}}$,$^{3,4}$
N. P. S. Mithun $^{\orcidlink{0000-0003-3431-6110}}$,$^{1}$ 
Santosh Vadawale $^{\orcidlink{0000-0002-2050-0913}}$,$^{1}$ 
\newauthor
Gulab Dewangan $^{\orcidlink{0000-0003-1589-2075}}$,$^{5}$
Poshak Gandhi $^{\orcidlink{0000-0003-3105-2615}}$,$^{6}$ 
\\ \\
$^{1}$Physical Research Laboratory, Navrangpura, Ahmedabad, Gujarat-380 009, India\\
$^{2}$Indian Institute of Technology Gandhinagar, Palaj, Gandhinagar, Gujarat-382 355, India \\
$^{3}$N\'{u}cleo de Astronom\'{i}a de la Facultad de Ingenier\'{i}a, Universidad Diego Portales, Av. Ej\'{e}rcito Libertador 441, Santiago, Chile \\
$^{4}$Kavli Institute for Astronomy and Astrophysics, Peking University, Beijing 100871, China \\
$^{5}$Inter-University Centre for Astronomy and Astrophysics (IUCAA), SPPU Campus, 411007, Pune, India\\
$^{6}$School of Physics \& Astronomy, University of Southampton, Highfield, Southampton SO17 1BJ, UK \\
}

\date{Accepted XXX. Received YYY; in original form ZZZ}

\pubyear{2023}

\begin{document}
\label{firstpage}
\pagerange{\pageref{firstpage}--\pageref{lastpage}}
\maketitle

\begin{abstract}
The circumnuclear material around Active Galactic Nuclei (AGN) is one of the essential components of the obscuration-based unification model. However, our understanding of the circumnuclear material in terms of its geometrical shape, structure and its dependence on accretion rate is still debated. 
In this paper, we present the multi-epoch broadband X-ray spectral modelling of a nearby Compton-thick AGN in Circinus galaxy. 
We utilise all the available hard X-ray ($>$ 10 keV) observations taken from different telescopes, {\ie} {\em BeppoSAX}, {\em Suzaku}, {\em NuSTAR} and {\em AstroSat}, at ten different epochs across 22 years from 1998 to 2020.
The 3.0$-$79 keV broadband X-ray spectral modelling using physically-motivated models, namely {\scshape MYTorus}, 
{\scshape borus02} and {\scshape UXCLUMPY}, infers the presence of a torus with a low covering 
factor of 0.28, an inclination angle of 77$^{\circ}$ $-$ 81$^{\circ}$ and Compton-thick line-of-sight
column densities ($N_{\rm H,LOS}$ = 4.13~$-$~9.26~$\times$~10$^{24}$~cm$^{-2}$) in all the epochs. The joint multi-epoch spectral modelling suggests that the overall structure
of the torus is likely to remain unchanged. However, we find tentative evidence for the variable line-of-sight column density on timescales ranging from one day to one week to a few years, suggesting a clumpy
circumnuclear material located at sub-parsec to tens of parsec scales. 
\end{abstract}




\section{Introduction}
\label{sec:intro}
The widely accepted unification model of Active Galactic Nuclei (AGN) invokes a gaseous and 
dusty axisymmetric toroidal structure popularly known as `torus' around the accreting 
Super-Massive Black Holes (SMBHs) \citep[e.g.,][]{Antonucci85,Urry95,Ramos17}. 
Based on the orientation of the obscuring torus, AGN can be classified mainly into two sub-classes named as type~1 (pole-on view) and type~2 (edge-on view).   
The evidence for the presence of the obscuring torus has come from a variety of observations that 
include the detection of broad emission lines in spectro-polarimetric observations of type 2 Seyferts 
\citep{Moran07}, biconical shapes of narrow-line regions (NLRs) in type~2 Seyferts \citep{Schmitt03}, 
and systematically higher X-ray absorbing column density in type~2 Seyferts \citep{Singh11,Ricci17}. 
To explain the observed differences between two sub-classes of Seyfert galaxies, most of the early studies assumed a doughnut-shaped uniform-density obscuring medium \citep[see][]{Netzer15}.     
Albeit, X-ray monitoring campaigns of nearby AGN ({\eg} NGC 1365; \citealp{Risaliti05}, NGC 4151; \citealp{Puccetti07}, and NGC 7582; \citealp{Rivers15}) finding the change in absorbing column density 
on timescales of days to several hours, favoured a clumpy obscuring medium. 
The discrete clouds forming the obscuring medium possibly exist at scales 
ranging from sub-parsec to a few hundred parsecs \citep{Bianchi12,Torricelli-Ciamponi14}.
Also, the column density of clumpy absorbing medium is likely to increase towards the equatorial plane 
\citep[see][]{Nenkova08a}. 
However, despite a large number of studies, the location, geometry, and physical state of the obscuring material are still widely debated \citep[e.g.,][]{Honig19,Saha22}.
\par
The high$-$resolution infrared (IR) observations from the Very Large Telescope Interferometer (VLTI) 
have revealed that, 
contrary to the prevalent paradigm of the classical torus, the dust around AGN is not distributed in one single toroidal structure \citep{Honig12,Tristram14,Leftley18}.
In fact, a two-component structure composed of an equatorial thin disk and a polar-extended cone-like feature, is implied from the modelling of IR interferometry data. 
However, IR observations probe only the dusty phase of the obscuring matter around the AGN and may not yield a complete picture of circumnuclear material.
The X-ray observations of AGN can provide insights into the geometry and distribution of circumnuclear material owing to the fact that X-ray emission emanating from the inner regions of the accreting system carries imprints of absorption and scattering caused by the gaseous phase of circumnuclear material \citep{Risaliti05, Morgan12}. 
\par 
The multi-epoch X-ray observations have revealed a significant change in the 
line-of-sight column density ($N_{\rm H,LOS}$) even in 
Compton-thick ($N_{\rm H,LOS}$ $>$ 1.5 $\times$ 10$^{24}$~cm$^{2}$) AGN (CT$-$AGN) 
({\eg} MRK~3; \citealp{Guainazzi16}, NGC~1068; \citealp{Zaino20}, NGC~1358; \citealp{Marchesi22}). 
The timescale for a significant variability in $N_{\rm H,LOS}$ depends on the distances, velocities and filling factors of the obscuring clouds around the SMBH. 
The type~2 AGN, in which line-of-sight passes through the obscuring torus, can be suitable targets 
to probe changes in $N_{\rm H,LOS}$ \citep{Ricci22}.
However, only a small number of such sources have been studied, hitherto, due to the unavailability of multi-epoch hard X-ray observations on timescales ranging from a few days to years.      
The hard X-ray ($>$ 10~keV) observations are crucial to probe any variation in the covering factor and to break
well-known line-of-sight column density and photon index ($N_{\rm H,LOS}$ $-$ $\Gamma$) degeneracy \citep[see][]{Puccetti14,Marchesi19}. 
Also, hard X-ray observations are useful in mitigating the influences of off-nuclear X-ray sources. 
With the availability of multi-epoch hard X-ray observations on timescales ranging from days to years, Circinus 
galaxy (hereafter `Circinus') is a suitable target for investigating changes associated with the AGN and reprocessing circumnuclear material. Due to its proximity (redshift ($z$) $0.00145\pm0.00001$ and
luminosity distance (D$_{\rm L}$) 4.2$\pm$0.8 Mpc; \cite{Freeman77}) and brightness, Circinus offers an advantage for performing an in-depth study of circumnuclear reprocessing material.
\par
In this paper, we investigate changes associated with the circumnuclear X-ray reprocessing material such as $N_{\rm H,LOS}$, average column density and covering factor in Circinus, for the first time, using multi-epoch hard 
X-ray observations from the {\em BeppoSAX}, {\em Suzaku}, Nuclear Spectroscopic Telescope Array
({\em NuSTAR}), and {\em AstroSat}
taken at ten different epochs across the period of 22 years from 1998 to 2020. 
The hard X-ray observations of the last three epochs {\ie}~2016 August 23, 2020 January 28 and 2020 November 26, 
are presented here for the first time.  
The paper is organized as follows.
In Section~\ref{sec:PreviousStudies}, we provide a brief description of previous studies performed on 
Circinus with the main focus on the results obtained from the hard X-ray observations.
In Section~\ref{sec:obs}, we give the details of X-ray observations and data reduction. 
In Section~\ref{sec:OffSrcs}, we present the X-ray spectral modelling of off-nuclear sources and contamination 
model. 
In Section~\ref{sec:BBmodels}, we describe multi-epoch broadband X-ray modelling. 
Section~\ref{sec:Discussion} is devoted to the discussion of plausible geometry, changes in line-of-sight 
column density and the location of obscuring clouds. In Section~\ref{sec:Summary}, we summarise the results of our study. 
\par
In our paper, we assume a flat $\Lambda$-cold dark matter cosmology with H$_{0}$ = 70 km s$^{-1}$ Mpc$^{-1}$, ${\Omega}_{\Lambda}$ = 0.73, and ${\Omega}_{\rm m}$ = 0.27, the same as those used in XSPEC 12.11.1c \citep{Arnaud96}. With these cosmological parameters, 1$^{\prime\prime}$.0 corresponds to 29 parsec at the distance of Circinus. We used Galactic neutral column density of 
$N_{\rm H}^{\rm gal}$ = 5.6 $\times$ 10$^{21}$ cm$^{-2}$ \citep{Kalberla05} toward the direction of Circinus. The errors quoted on the spectral parameters are of 90 per cent confidence, unless stated otherwise.
\section{Previous studies on Circinus}
\label{sec:PreviousStudies}
Based on the optical spectroscopic observations, Circinus is classified as a Seyfert type 2 AGN 
with starburst activity \citep{Freeman77,For12}.  
The mass of SMBH ($M_{\rm BH}$) in the Circinus is estimated to be 1.7 $\times$ 10$^{6}$ $M_\odot$ \citep{Koss17} using the relationship between black hole mass and stellar velocity dispersion \citep{Kormendy13}. 
The bolometric luminosity is estimated to be $L_{\rm bol}$ = 4 $\times$ 10$^{43}$ erg s$^{-1}$ from the 
mid-IR (MIR) nuclear spectrum \citep{Moorwood96}. 
The estimates of black hole mass and bolometric luminosity suggest an Eddington ratio ($L/L_{\rm Edd}$) of 0.2.
The high$-$resolution VLTI mid-infrared interferometric instrument (MIDI) observations revealed that the parsec-scale MIR emission around the AGN in the Circinus is composed of two distinct components : 
(i) a disc-like component coinciding with the disc observed 
in maser emission, and (ii) a component extending in a polar direction, along the 
ionisation cone seen in the optical \citep{Tristram07, Tristram14}. 
The polar component is found to be responsible for up to $\sim 80\%$ of the MIR emission on parsec-scales \citep{Tristram14}.
\par 
Circinus is widely studied in the X-ray wavelengths and is classified as a CT$-$AGN.
The X-ray observations below 10 keV from the 
{\em ASCA}, {\em Chandra}, and {\em XMM-Newton} showed only a reflection-dominated spectrum with a prominent
Fe K$\alpha$ emission line \citep[see][]{Matt96,Guainazzi99,Marinucci13}.
Early observations at hard X-ray from the {\em BeppoSAX} showed a high absorbing column density of 
$N_{\rm H}$ $\sim$ 4 $\times$ 10$^{24}$ cm$^{-2}$, confirming Circinus to be a CT-AGN.
Interestingly, two {\em BeppoSAX} observations performed almost three years apart during 1998 and 2001 reported a dramatic flux ($\sim$ 50 per cent) and spectral variation \citep{Bianchi02a}. 
The observed variability was ascribed to an extremely variable ultra$-$luminous X-ray (ULX) source named 
Circinus Galaxy X1\citep[CGX1;][]{Bauer01} that contaminated {\em BeppoSAX} spectrum obtained with 2$^{\prime}$.0 
extraction region. However, an intrinsic variation in AGN flux or line-of-sight column density could not be completely 
ruled out.
Using 3$-$70 keV {\em Suzaku} observations, \cite{Yang09} reported the presence of a direct AGN 
component transmitted through a column density of $\simeq 4 \times 10^{24}$ cm$^{-2}$. 
Later on, using {\em Chandra} and {\em XMM-Newton} observations of higher spatial resolution 
\cite{Arevalo14} accounted for the contribution from the off-nuclear contaminating X-ray sources, and reported that the hard X-ray spectrum is Compton-scattered by an optically-thick torus having equatorial column density 6 $-$ 10 $\times$ 10$^{24}$ cm$^{-2}$ with a relatively steeper photon index 
of $\Gamma$ = 2.2 $-$ 2.4. Notably, {\em NuSTAR} observations did not support the detection of 
transmitted AGN component, which was earlier suggested with the {\em Suzaku} observations. 
More recently, \cite{Andonie22} showed that the 3$-$70 keV {\em NuSTAR} spectrum of Circinus 
could also be fitted with a model considering the reprocessing components consisted of an accretion disc, BLR, a 
flared disc and a hollow cone in the polar direction. 
\section{X-ray observations and Data reduction}
\label{sec:obs}
For our study, we utilised all the available hard X-ray observations 
($E$ $>$ 10 keV) taken from various telescopes ({\eg} {\em BeppoSAX}, {\em Suzaku} {\em NuSTAR}
and {\em AstroSat}) during 1998 to 2020. Circinus is observed with the {\em NuSTAR} at six different epochs, twice with the {\em BeppoSAX}, and once each with the {\em AstroSat} and {\em Suzaku}.
To assess the contamination from the off-nuclear X-ray
sources, we utilised {\em Chandra} and {\em XMM-Newton} observations. 
The {\em XMM-Newton} observations taken on 2013 February 03 and 2016 August 23 were performed 
quasi$-$simultaneously with the {\em NuSTAR} observations of corresponding epochs. 
The start and end times of each {\em XMM-Newton} and {\em NuSTAR} pair of observations are within 12 hours. 
We list the basic details of all the X-ray observations in Table~\ref{tab:ObsLog}. 
In the following subsections, we describe our observations and data reduction procedures.    

\begin{table*}
\centering
\begin{adjustbox}{width=\textwidth}
\begin{threeparttable}
\caption{Summary of X-ray observations used in this work}
\begin{tabular}{lcccccccc}
\hline
Epoch & Instrument  & Date \&        &  ObsID        & Energy    & Detector &  $T_{\rm exp}$ & Count-  & Off-   \\
      &             &  start time    &               & band      &          &                & rate   & axis   \\ 
      &                &             &               &  (keV)    &          & (ks)           & (cts~s$^{-1}$) &  ($^{\prime}$) \\
(1)   & (2)            &  (3)        & (4)           &   (5)     & (6)      &  (7)    & (8)   &  (9)  \\ \hline
 1    & {\em NuSTAR} & 2020-11-26T05:51:09  & 80601502001  & 3$-$79 & FPMA  & 106.1 & 0.86$\pm$0.01    &  2.65  \\
      &              &             &              & 3$-$79   & FPMB  & 105.2 & 0.82$\pm$0.01   & 2.65  \\
 2    & {\em AstroSat} & 2020-01-28T21:34:57 & A07\_100T02\_9000003470 & \multirow{2}{*}{0.3$-$7.0} & \multirow{2}{*}{SXT}   & \multirow{2}{*}{57.5}  & \multirow{2}{*}{0.13$\pm$0.01}    &  0.06   \\
      &                &             & A07\_100T02\_9000003476 &         &   &      &        &      \\
      &  &  &  & 3$-$80  & LAXPC  & 85.4 & 6.72$\pm$0.10   & 0.06    \\
      &  &  &  & 22$-$70 & CZTI  & 71.0 & 0.56$\pm$0.06    &  ...  \\
      & {\em XMM~Newton} & 2018-09-18T13:48:49 & 0824450301 &  0.5$-$10 & pn &  85.5 &  1.09$\pm$0.01 &  6.04  \\

3     & {\em NuSTAR}  & 2016-08-23T06:41:08 & 90201034002  & 3$-$79   & FPMA  & 49.8 & 0.57$\pm$0.01   &  5.76  \\
      &               &            &              &  3$-$79    & FPMB  & 49.7 & 0.55$\pm$0.01   & 5.76   \\
      & {\em XMM~Newton} & 2016-08-23T16:53:33 & 0792382701  & 0.5$-$10 &  pn   & 16.2 & 1.72$\pm$0.01 &  5.78  \\
4     & {\em NuSTAR} & 2013-02-05T05:06:07 & 30002038006  & 3$-$79 & FPMA    & 36.2    & 0.82$\pm$0.01 &  3.05     \\
      &              &             &              & 3$-$79 & FPMB    & 36.1    & 0.85$\pm$0.01 &  3.05           \\
5     & {\em NuSTAR} & 2013-02-03T03:01:07 & 30002038004  & 3$-$79 & FPMA    & 40.3 & 0.84$\pm$0.01           &  2.91  \\  
      &              &             &              & 3$-$79 & FPMB    & 40.2 & 0.86$\pm$0.01          & 2.91   \\
      & {\em XMM~Newton}       & 2013-02-03T07:24:11 & 0701981001  & 0.5$-$10 &  pn   & 36.2 & 1.89$\pm$0.01  &   3.69 \\
6     & {\em NuSTAR} & 2013-02-02T01:01:07 & 30002038002  & 3$-$79 & FPMA    & 18.3   & 0.92$\pm$0.01     &  2.85   \\  
      &              &             &              & 3$-$79 & FPMB    & 18.3  & 0.92$\pm$0.01     &  2.85  \\ 
7     & {\em NuSTAR} & 2013-01-25T03:51:07 & 60002039002  & 3$-$79 & FPMA    & 53.9   & 1.12$\pm$0.01 &  2.73   \\
      &              &             &              & 3$-$79 & FPMB    & 53.8  & 1.05$\pm$0.01  &   2.73  \\
      & {\em Chandra} & 2010-12-17T18:10:27 & 12823        & 0.5$-$10 & ACIS-S & 152.4 & 0.06$\pm$0.01 &  0.21  \\
8     & {\em Suzaku}   & 2006-07-21T12:29:57  & 701036010  & 0.5$-$10 & XIS-0  & 108.0   & 0.54$\pm$0.01 & 0.93  \\
&                      &             &            & 0.5$-$10 & XIS-1  & 108.0   & 0.53$\pm$0.01 & 0.76   \\
&                      &             &            & 0.5$-$10 & XIS-2  & 108.0   & 0.53$\pm$0.01 & 0.97   \\
&                      &             &            & 0.5$-$10 & XIS-3  & 108.0   & 0.52$\pm$0.01 & 1.09   \\
&                      &             &            & 10$-$70  & HXD-PIN & 88.3 & 0.40$\pm$0.01   & 3.91   \\
&                      &             &            & 50$-$120 & HXD-GSO & 88.3 & 0.14$\pm$0.01   &  3.91  \\
9     & {\em BeppoSAX}   & 2001-01-07T06:36:41 & 5114000100 & 2.0$-$10 & MECS    & 51.7 & 0.18$\pm$0.01   &  1.77  \\
      &                  &            &            & 15$-$100 &  PDS   &   37.9 & 1.72$\pm$0.03   & ...    \\
      & {\em XMM-Newton} & 2001-08-06T08:54:51 & 0111240101 & 0.5$-$10 &  pn   & 103.0   & 2.45$\pm$0.01     &  1.71   \\
10    & {\em BeppoSAX}   & 1998-03-13T06:32:49 & 5004700200 & 2.0$-$10 & MECS  & 71.5 & 0.14$\pm$0.01 &  1.88  \\
      &                  &            &            & 15$-$100 &  PDS  & 63.3 & 1.83$\pm$0.03     &   ... \\
\hline
\end{tabular}
\label{tab:ObsLog}
\begin{tablenotes}
\normalsize
\item Notes -  $T_{\rm exp}$ is the net exposure time after removal of bad time$-$intervals. 
The net count rate in the total energy band is estimated after the removal 
of bad time$-$intervals and subtraction of background. 
The parameters of {\em AstroSat} observations are combination of observations taken under the two different observation IDs. 
Epochs are listed in chronological order based on the availability of hard X-ray ($>$ 10~keV) observations. 
\end{tablenotes}
\end{threeparttable}
\end{adjustbox}
\end{table*}
\subsection{{\em NuSTAR}}
\label{sec:NuSTAR}
Circinus has been observed with the {\em NuSTAR} \citep{Harrison13} with its two co-aligned units
having focal plane modules FPMA and FPMB.
We note that all the {\em NuSTAR} observations, except one taken on 2013 January 25 
have targeted an ULX binary named ULX5 \citep[see][]{Mondal21} residing 4$^{\prime}$ to the south$-$west of the AGN, thus imaged Circinus at off-axis. 
The {\em NuSTAR} with 12$^{\prime}$.5 $\times$ 12$^{\prime}$.5 ﬁeld-of-view (FOV) and 
angular resolutions of 18$^{\prime\prime}$ Full Width at Half Maximum (FWHM) offers a clear detection of 
Circinus, along with the ULX5. 
\par 
The {\em NuSTAR} data were reduced using the standard pipeline ({\ttfamily NUPIPELINE}) 
provided in the {\em NuSTAR} Data Analysis Software ({\ttfamily NUSTARDAS}, v2.1.2) 
within the HEASoft package (v6.30), in combination with the calibration database (v20220706). 
The unﬁltered level 1 event lists were screened to 
reduce the internal background at high energies via standard depth corrections and 
removal of South Atlantic Anomaly (SAA) passages. 
The latest calibration files were used to create level~2 event files.    
We extracted the source spectra and lightcurves of AGN by using an aperture with a radius of
100$^{\prime\prime}$ for both FPMA and FPMB. The background spectra were extracted
from a source-free region in the same chip using a circular aperture of the same size.
We note that the 2016 {\em NuSTAR} observations show Circinus at a large off-axis angle of 5$^{\prime}$.7 with an elongated PSF. Hence, to include all the source emission with minimum contamination from the background, we extracted the source spectrum from an elliptical region (major axis = 90$^{\prime\prime}$, minor axis = 50$^{\prime\prime}$ 
and PA = 44$^{\circ}$).
The net exposure times and count rates for both FPMA and FPMB of our observations are listed in 
Table~\ref{tab:ObsLog}. 
To apply $\chi^2$ statistics, we binned our spectra to have a minimum of 50$-$100 counts per bin.
\subsection{{\em XMM-Newton}}
\label{sec:XMM-Newton}
To study the broadband spectral properties of Circinus, we used {\em XMM-Newton} observations 
taken simultaneously with the {\em NuSTAR} observations on 2013 February 03 and 2016 August 23, respectively 
(see Table~\ref{tab:ObsLog}). Both of these observations were aimed at ULX5. 
The AGN located nearly 4$^{\prime}$ away is 
well detected due to sufficiently high angular resolution (6$^{\prime\prime}$.0) and 
large FOV (30$^{\prime}$) of EPIC pn/MOS cameras. 
Also, to check variability in off-nuclear sources, we used {\em XMM-Newton} observations taken on
four different epochs, {\em i.e,} 2018 September 18, 2016 August 23, 2013 February 03 and 2001 August 06.
During all the epochs, EPIC pn observations were taken in the full window mode with the medium filter,
except for 2016 when the thin filter was used.  
\par
We reduced the data using Science Analysis System (SAS v19.0.0) 
following the standard procedure. 
We considered only single and double events with quality ﬂag set equal to 0. 
We obtained calibrated event files by using the latest calibration files. 
The good time interval event files were generated by removing time intervals of 
flaring background with count rates exceeding 0.8 ct s$^{-1}$ for pn in the 10$-$12 keV band. 
%
The net exposure times for pn in the 0.5$-$10 keV energy band are found to 
be 85.5~ks, 16.2~ks, 36.2~ks and 103.0~ks for 2018, 2016, 2013 and 2001 observations, respectively (see Table~\ref{tab:ObsLog}). 
We generated response matrices and ancillary response files using the {\tt rmfgen} and 
{\tt arfgen} tasks, respectively. The spectra were binned to have a minimum of 30 counts per bin.
\par
To perform the joint {\em XMM-Newton} and {\em NuSTAR} spectral fit, we extracted the {\em XMM-Newton} pn spectrum using
100$^{\prime\prime}$ radius to match the extraction region of the {\em NuSTAR} spectrum.
The background spectra were extracted using a polygon region covering an area devoid of any source emission in the same chip.   
We point out that the EPIC pn image, with an angular resolution of 6$^{\prime\prime}$.0, enables us to spatially resolve AGN and neighbouring off-nuclear X-ray sources CGX1 and CGX2.
We extracted CGX1 and CGX2 spectra using circular apertures of radii 7$^{\prime\prime}$.5 and 11$^{\prime\prime}$,
respectively.
\subsection{{\em AstroSat}}
\label{sec:AstroSAT}
We (PI: A. Kayal) observed Circinus using {\em AstroSat} during 2020 January 28 to 2020 February 01.
The Soft X-ray Telescope (SXT) onboard {\em AstroSat} was kept as the prime instrument for these observations.
\subsubsection{SXT} 
\label{sec:SXT}
The SXT, a grazing incidence X-ray telescope operating in 0.3--7.1 keV band,
offers an angular resolution of 2$^{\prime}$.0 \citep{Singh17}.
The SXT observations of Circinus were performed in photon counting (PC) mode. 
We reduced data using {\tt SXTPIPELINE} (version 1.4b\footnote{https://www.tifr.res.in/~astrosat_sxt/sxtpipeline.html}),
and generated cleaned and calibrated level 2 event files for each orbit.
The data reduction pipeline includes standard processes such as event extraction,
screening criteria ({\eg} eliminating bad pixels, SAA passage, and events with grades $>$12), 
and calibration. The cleaned and calibrated level 2 event files for each orbit were then merged 
using {\tt SXTPYJULIAMERGER\_v02} that removed any overlap between the consecutive orbits. 
The merged level 2 event file gives an effective exposure time of $57.49$ ks. 
The scientific data products, {\eg} images, light curves and spectra, were extracted using
the {\tt XSELECT} task within the HEASoft package.
We extracted the source spectrum using a circular region of radius 13$^{\prime}$ with encircled energy fraction of $92.3\%$. 
For spectral analysis, we used the SXT spectrum along with the relevant response files and background spectrum provided by the SXT instrument team.
Considering calibration uncertainties in the response, we added a systematic uncertainty of 3 per cent to
the SXT data \citep[{\eg}][]{Jithesh19,Swain23}.
To account for the slight change in the SXT gain \citep{Singh17} we used gain fit task in the XSPEC by fixing the
slope to 1 keV and varying the offset parameter. Subsequently, we fixed the offset parameter to the resultant
value of 0.03 keV.
\subsubsection{LAXPC} 
\label{sec:LAXPC}
The LAXPC, sensitive in the 3.0--80 keV band, is a non-focusing instrument \citep{Antia17}.
The {\em AstroSat} houses three identical units of proportional counters (LAXPC10, LAXPC20 and LAXPC30), filled with highly pressurised xenon gas.
The LAXPC30 unit is suspected to have undergone a gas leakage resulting in a continuous gain shift and LAXPC10 is unstable \citep[see][]{Antia17}. Hence, we used data from the LAXPC20 unit only.
The data from LAXPC were processed and analysed using the latest version (August 15, 2022) of the LAXPC pipeline package
{\tt laxpcSoft}, provided by the LAXPC POC\footnote{https://www.tifr.res.in/astrosat{\_}laxpc/software.html}. 
The level~2 event file was generated by combining the level~1 event files from all orbits and 
removing any overlap between two consecutive orbits. We obtain a net exposure time of 85.4 ks 
by considering only good time intervals and selecting events from only the top layer of the LAXPC20 unit. 
The spectrum and lightcurve were extracted from the level 2 file by applying appropriate response functions and gain variations.
We applied gain shift utility to account for the shift in 
gain values of the background spectrum during the time of observations.
We obtained a gain offset of $-0.4$~keV, a value similar to the one reported in previous studies
\citep[see][]{Antia21}.
Due to the completely background$-$dominated spectrum at higher energies, we used the LAXPC spectrum only in the 4$-$20 keV energy band. 
We also added systematic uncertainty of 3.0 per cent to account for calibration uncertainties in the response
\citep[see][]{Antia21}.
\subsubsection{CZTI} 
\label{sec:CZTI}
The Cadmium Zinc Telluride Imager \citep[CZTI,][]{Bhalerao17} onboard {\em AstroSat} is a hard X-ray 
coded mask instrument operating in the 22$-$200 keV energy range. 
It contains four identical but independent quadrants, with each quadrant consists of sixteen CZT detector modules.
We reduced data using CZTI data analysis pipeline\footnote{http://astrosat-ssc.iucaa.in/cztiData} version 3.0.  
From the raw event list, we generated cleaned event files by applying the
recommended good time interval (GTI) selection criteria. We generated background subtracted spectra for each quadrant 
by using the \verb|cztbindata| task. The spectra of all four quadrants from two sets of observations 
(A07\_100T02\_9000003470 and A07\_100T02\_9000003476) were 
added together by using the \verb|cztaddspec| task. The combination of two observations provided 
a net exposure time of 71.0 ks. The source counts were detected above 5$\sigma$ in the 22$-$70 keV energy range.
Hence, we used the 22$-$70 keV CZTI spectrum after grouping it into broader energy bins.
\subsection{{\em Suzaku}}
The Circinus was observed with the {\em Suzaku} \citep{Mitsuda07} on 2006 July 21, for nearly 140 ks.
There are four X-ray Imaging Spectrometer \citep[XIS,][]{Koyama07} CCDs 
located at the focal planes of the respective foil mirrors X-ray telescopes,
and a non-imaging collimated Hard X-ray Detector (HXD) onboard {\em Suzaku}. 
The four XIS CCDs, {\ie} XIS0, XIS1, XIS2, and XIS3 are sensitive in 0.4$-$10 keV energy band.
The three XIS units have front-illuminated CCDs, while XIS1 with back-illuminated CCDs provides better 
quantum efficiency in the sub-keV energy range.   
The HXD uses 16 phoswich counter detectors, with each unit consisting
of a GSO scintillation counter and PIN 
silicon diodes. The PIN detector is sensitive in the energy range of 12$-$60 keV, while GSO is sensitive above 
40 keV. 
\par 
We reduced the XIS and HXD data using the HEASoft software package (version 6.30) and following the steps given in
Suzaku Data Reduction Guide\footnote{https://heasarc.gsfc.nasa.gov/docs/suzaku/analysis/abc/}.
From cleaned calibrated event files, we extracted XIS spectra using a circular extraction region with a radius 
of 2$^{\prime}$.5 centered at the Circinus. The background spectra were extracted from a source-free region. 
The HXD spectra were generated using the {\tt HXDPINXBPI} script. 
The HXD being a collimating instrument, requires background estimation from the non-X-ray instrumental 
background (NXB) and cosmic X-ray background (CXB). We utilised the response and NXB files provided
by the Suzaku team. To apply ${\chi}^{2}$ minimisation, we binned PIN and GSO spectra such that each bin has a 
minimum signal-to-noise ratio of 3. The binned spectra of PIN and GSO have energy ranges of 10$-$70 keV and 
50$-$120 keV, respectively. The count rates for PIN and GSO are $0.40\pm0.01$ ct s$^{-1}$ and $0.14\pm0.01$
ct s$^{-1}$, respectively. 
We note that, due to relatively coarse angular resolution, both XIS as well as HXD spectra of Circinus are 
contaminated by the neighbouring off-nuclear sources.  
\subsection{{\em BeppoSAX}}
The X-ray satellite {\em BeppoSAX} \citep{Boella97} carried four co$-$aligned instruments, namely 
a Low Energy Concentrator Spectrometer (LECS), three Medium Energy Concentrator Spectrometers (MECS), a High$-$Pressure Gas Scintillation Proportional Counter (HPGSPC), and a Phoswich Detector System (PDS). 
With imaging capabilities, LECS and MECS operated in 0.1$-$10 keV and 1.3$-$10 keV energy ranges, respectively
and both have an angular resolution of nearly 1$^{\prime}$.2 arcmin at 6 keV.
The HPGSPC and PDS are collimating instruments and cover 4$-$120 keV and 15$-$200 keV energy
ranges, respectively. Since PDS is more sensitive than HPGSPC in the overlapping energy range, we preferred to 
use PDS data. Also, considering the better sensitivity of MECS than LECS in the overlapping band,
we used the MECS spectrum. Thus, our broadband spectral analysis is based on MECS and PDS spectra.
We obtained MECS and PDS spectral products from the SSDC multi-mission interactive 
archive\footnote{https://www.ssdc.asi.it/mmia/index.php?mission=saxnfi}.  
The corresponding response and background files were taken from the CALDB directory. 
\subsection{{\em Chandra} ACIS-S}
\begin{figure*}
\centering
\includegraphics[angle=0, trim={0 0.0cm 0 0}, clip]{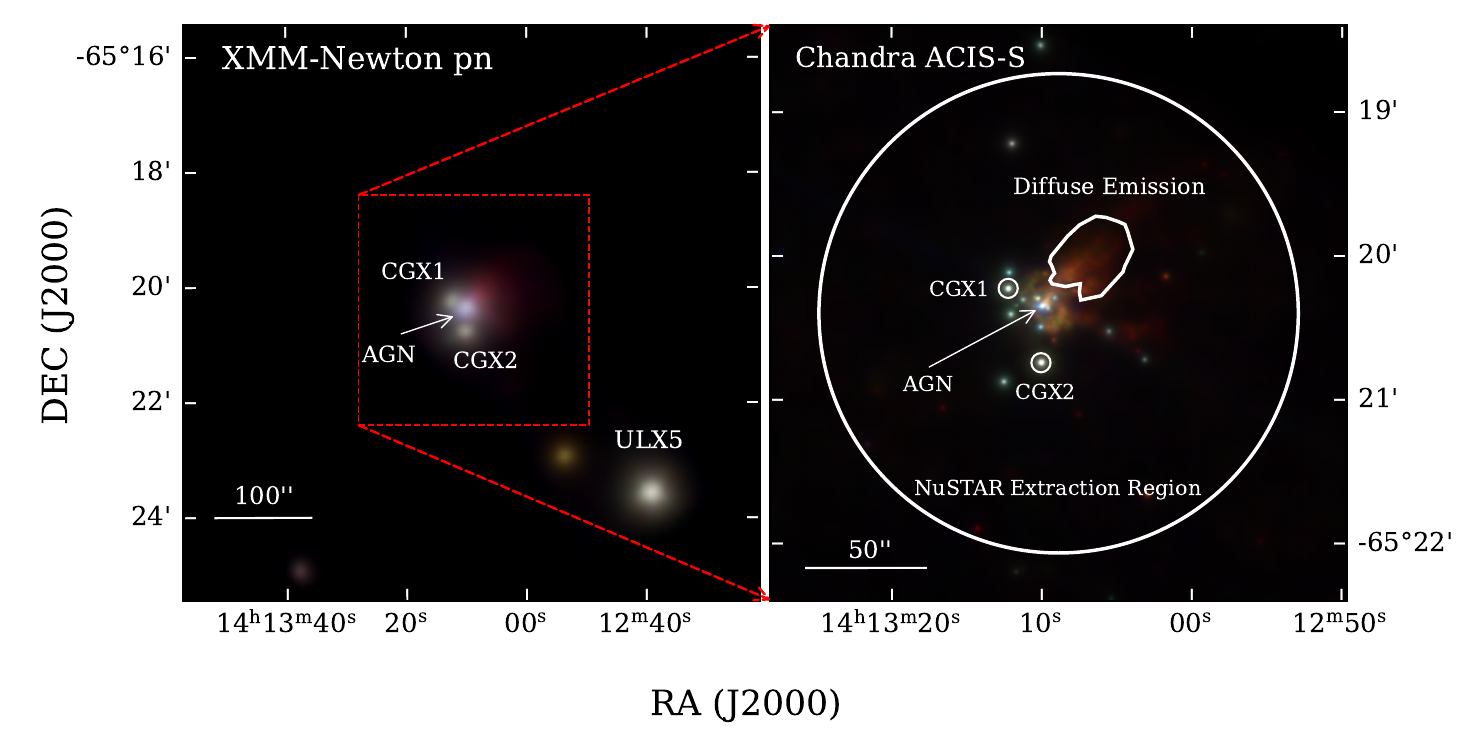}
\caption{{\it Left panel} : The {\em XMM-Newton} pn image of Circinus in which the AGN and the off-nuclear  contaminating sources CGX1, CGX2 and ULX5 are marked. {\it Right panel} : The zoom-in view of the central
region as seen in the sensitive {\em Chandra} ACIS-S image of higher resolution obtained from the longest
exposure time of $152$ ks.
The extraction regions of CGX1, CGX2 and extended diffuse emission, are marked.
The location of the AGN is indicated by an arrow. The large circle represents the {\em NuSTAR} extraction region of
100$^{\prime\prime}$ radius.
Three different colours in both images represent three different energy bands {\ie} 0.3$-$1.5 keV (red), 1.5$-$2.5~keV (green) and 2.5$-$8.0~keV (blue).}
\label{fig:ChandraIm}
\end{figure*}

The high spatial resolution (0$^{\prime\prime}$.5) of {\em Chandra} ACIS-S enables us to resolve and assess
the contribution of off-nuclear sources (see Figure~\ref{fig:ChandraIm}). 
To perform spectral analysis of off-nuclear sources and extended diffuse emission, we 
used {\em Chandra} ACIS-S observations of Circinus taken on 17 December 2010 (see Section~\ref{sec:OffSrcs}). 
With 152 ks exposure time, these are the deepest imaging observations performed with the {\em Chandra}. 
The ACIS-S, sensitive in the 0.4$-$8.0 keV energy range, consists of four front-illuminated and
two back-illuminated CCDs.
We reduced {\em Chandra} data using the CIAO software\footnote{https://cxc.harvard.edu/ciao/} (v4.14) and
the latest available calibration files from CALDB v4.9.8. In data processing, we 
removed 0$^{\prime\prime}$.5 pixel randomization, and corrected for charge transfer inefficiency 
(CTI), excluded bad pixels and time intervals of high background. 
With a 3.2s frame time, no source except AGN suffers from the pile-up. 
Using calibrated and cleaned event file, we extracted spectra of off-nuclear sources. 
\section{Assessment of contamination from off-nuclear sources} 
\label{sec:OffSrcs}
We point out that {\em Chandra} ACIS and {\em XMM-Newton} pn images offering higher angular 
resolutions of 0$^{\prime\prime}$.5 and 6$^{\prime\prime}$.0, respectively, show 
off-nuclear X-ray sources CGX1 and CGX2 located close to the AGN (see Figure~\ref{fig:ChandraIm}).  
CGX1 is located north$-$east to the AGN at a distance of 15$^{\prime\prime}$, while CGX2 is detected at 
a distance of 25$^{\prime\prime}$ towards the south of AGN. 
The {\em NuSTAR} images with an angular resolution of 18$^{\prime\prime}$ (FWHM) are unable 
to spatially resolve CGX1 and CGX2. 
Therefore, the {\em NuSTAR} spectrum includes contributions from these two neighbouring off-nuclear X-ray sources 
in addition to that from AGN. 
We note that the deep {\em Chandra} ACIS image taken in 2010 December shows several faint off-nuclear sources in addition to CGX1 and CGX2. 
Although, \cite{Arevalo14} demonstrated that the cumulative contribution 
from the faint sources is insignificant in comparison to the CGX1 and CGX2, which is further vindicated by the fact that the
{\em XMM-Newton} pn images of different epochs detect only relatively bright CGX1 and CGX2. 
Hence, we account for the contamination from CGX1 and CGX2 while analysing the {\em NuSTAR} spectra.
We recall that the {\em XMM-Newton} and {\em NuSTAR} images show ULX5 \citep{Walton13} as a bright X-ray source
located 4$^{\prime}$.5 away south-west to the AGN. The {\em BeppoSAX}, {\em Suzaku} and {\em AstroSat} observations
with a fairly coarse angular resolution of collimating instruments are  unable to resolve AGN 
and neighbouring sources CGX1, CGX2 as well as ULX5. Except for {\em Suzaku} XISs spectra, 
the soft X-ray spectra from {\em BeppoSAX} MECS and {\em AstroSat} SXT include the contribution from the ULX5.
Therefore, while modelling the broadband spectra of these instruments, we account for the contamination from the
ULX5 too, in addition to the CGX1 and CGX2.
\par
To assess and remove the contribution of contaminating off-nuclear X-ray sources,  
we followed a procedure similar to that presented by \cite{Arevalo14}. 
Using {\em Chandra} ACIS and {\em XMM-Newton} pn observations, we model 0.5$-$10 keV X-ray spectra 
of individual contaminants, {\ie} CGX1, CGX2, ULX5 and extended emission.     
For each contaminating source, we attempt to achieve a baseline model that can provide a 
reasonably good fit to the spectra of all epochs. 
With the knowledge of the spectral shape and parameters of contaminants, we formulate a contamination model that accounts for the contribution of all contaminants. 
In the following subsections, we discuss spectral modelling of the contaminating sources and formulation of the contamination model.
\begin{table*}
\centering
\renewcommand{\arraystretch}{1.2}
\begin{threeparttable}
\caption{The best$-$fitted spectral parameters and fluxes of CGX1, CGX2 and ULX5 during different epochs}
\begin{tabular}{ccccccc}
\hline
Parameters & Unit & 2001 & 2010  & 2013 & 2016  &  2018 \\
(1)              & (2 ) & (3)  &  (4)    & (5)  &  (6)  &  (7)     \\
\hline
\multicolumn{7}{c}{{\bf CGX1}~[Model: {\scshape tbabs $\times$ tbabs $\times$ po}]}  \\
$N_{\rm H}^{\rm gal}$ & $10^{22}$ cm$^{-2}$ & 0.56$^{f}$ & 0.56$^{f}$ & 0.56$^{f}$ & 0.56$^{f}$ & 0.56$^{f}$ \\
$N_{\rm H}$ & $10^{22}$ cm$^{-2}$ & 0.69$^{+ 0.03}_{- 0.03}$ &  0.51$^{+ 0.05}_{- 0.05}$ & 0.32$^{+ 0.04}_{- 0.04}$ & 0.42$^{+ 0.06}_{- 0.06}$ & 0.35$^{+ 0.05}_{- 0.05}$ \\
$\Gamma$ &       & 1.8$^f$  &  1.80$^{+0.06}_{-0.06}$ & 1.8$^f$  & 1.8$^f$  & 1.8$^f$ \\
$Norm_{\Gamma}$ & 10$^{-4}$ & $1.10^{+ 0.22}_{- 0.22}$ &  2.48$^{+ 0.21}_{- 0.19}$ & $3.96^{+ 0.16}_{- 0.16}$ & $5.98^{+ 0.30}_{- 0.30}$ & $1.87^{+ 0.08}_{- 0.08}$ \\
$F_{\rm 0.5 - 10.0~keV}$ & 10$^{-12}$~erg cm$^{-2}$ s$^{-1}$ & $3.86^{+ 0.06}_{- 0.06}$ &  0.97$^{+ 0.03}_{- 0.03}$ & $1.50^{+ 0.05}_{- 0.05}$ & $2.21^{+ 0.10}_{- 0.10}$ & $0.70^{+ 0.26}_{-0.28}$  \\
$\chi^2_{\rm r}$ (dof) & & $1.28~(647)$ & 1.09 (299) & $1.06~(191)$ & $0.69~(116)$ & $1.11 ~(202)$ \\ \hline
\multicolumn{7}{c}{{\bf CGX2}~[Model: {\scshape tbabs $\times$ (tbabs $\times$ vpshock + tbabs $\times$ vpshock)}]}\\
$N_{\rm H}^{\rm gal}$ & $10^{22}$ cm$^{-2}$ & 0.56$^{f}$ & 0.56$^{f}$ & 0.56$^{f}$ & 0.56$^{f}$ & 0.56$^{f}$ \\
$N_{\rm H}$ & $10^{22}$ cm$^{-2}$ & $0.70^{+0.08}_{-0.08}$ & $0.44^{+ 0.04}_{-0.04}$ & $0.30^{+ 0.10}_{- 0.09}$ & $0.35^{+ 0.18}_{-0.19}$ & $0.23^{+ 0.08}_{- 0.08}$ \\
$kT^{\rm R}$ & keV & $8.0^f$ &  $8.0^f$  & $7.89^{+ 0.75}_{- 0.69}$ & $8.0^f$ & $8.0^f$ \\
$kT^{\rm F}$ & keV & $3.0^f$ &   $3.0^f$ & $2.98^{+ 1.81}_{- 0.87}$ & $3.0^f$ & $3.0^f$ \\
$\tau_{\rm u}^{\rm R}$ & $10^{13}$ s cm$^{-3}$ & $5.0^{\it f}$ & $5.0^{\it f}$ & $5.00^{\rm peg}_{- 4.58}$ & {$5.0^f$} & {$5.0^f$} \\
$\tau_{\rm u}^{\rm F}$ & $10^{11}$ s cm$^{-3}$ & $5.5^{\it f}$ & $5.5^{\it f}$ & $5.50^{+ 7.01}_{-2.47}$ & {$5.5^f$} & {$5.5^f$} \\
$Norm^{\rm R}$ & $10^{-4}$ & $6.81^{+ 0.40}_{- 0.41}$ & $7.13^{+0.33}_{-0.34}$ & $5.98^{+ 0.65}_{- 0.91}$ & $4.88^{+ 0.65}_{- 0.70}$ & $4.16^{+ 0.25}_{- 0.26}$ \\
$Norm^{\rm F}$ & $10^{-4}$ & $3.34^{+ 0.70}_{- 0.68}$ & $2.96^{+0.47}_{-0.46}$ & $2.15^{+ 0.86}_{- 0.77}$ & $0.99^{+ 1.03}_{- 0.90}$ & $0.64^{+ 0.37}_{- 0.33}$ \\
$F_{\rm 0.5-10~keV}$ & $10^{-12}$ erg cm$^{-2}$ s$^{-1}$ & $1.49^{+ 0.04}_{- 0.04}$ & $1.59^{+0.03}_{-0.03}$ & $1.34^{+ 0.04}_{- 0.04}$ & $1.00^{+ 0.08}_{- 0.08}$ & $0.85^{+ 0.03}_{- 0.03}$ \\
$\chi^2_{\rm r}$ (dof) &  & $0.96 (472)$ & $1.23~(338)$ & $0.95 (264)$ & $1.26~(87)$ & $1.13~(358)$ \\ \hline
\multicolumn{7}{c}{{\bf ULX5}~[Model: {\scshape tbabs $\times$ diskbb}]} \\
$N_{\rm H}^{\rm gal}$ & $10^{22}$ cm$^{-2}$ & $0.56^f$ &  & $0.56^f$ & $0.56^f$ & $0.56^f$ \\
$T_{\rm in}$ & keV & $1.17^{+ 0.03}_{- 0.03}$ &  & $1.85^{+ 0.02}_{- 0.02}$ & $1.08^{+ 0.03}_{- 0.03}$ & $1.78^{+ 0.01}_{- 0.01}$ \\
$Norm_{\rm diskbb}$ & $10^{-2}$ & $2.37^{+ 0.27}_{- 0.24}$ &  & $3.26^{+ 0.16}_{- 0.15}$ & $8.88^{+ 0.10}_{- 0.10}$ & $3.53^{+ 0.11}_{- 0.10}$ \\
$F_{\rm 0.5 - 10.0~keV}$ & $10^{-12}$ erg cm$^{-2}$ s$^{-1}$ & $0.63^{+ 0.01}_{- 0.02}$ &  & $6.07^{+ 0.07}_{- 0.08}$ & $1.66^{+ 0.04}_{- 0.06}$ & $5.88^{+ 0.04}_{- 0.05}$ \\
$\chi^2_{\rm r}$ (dof) &  & $1.75~(288)$ &  & $0.97~(896)$ & $1.06~(205)$ & $1.13~(1198)$ \\
\hline
\multicolumn{7}{c}{{\bf Diffuse emission}~[Model: {\scshape tbabs $\times$ apec + MYTorus}]} \\
$N_{\rm H}^{\rm gal}$ & $10^{22}$ cm$^{-2}$ &  & $0.56^f$ &  &  &  \\
$kT$ & keV &  & $0.82^{+ 0.02}_{- 0.02}$ &  &  &  \\
$Z$ & $Z_{\odot}$ &  & $0.19^{+ 0.02}_{- 0.02}$ &  &  &  \\
$Norm_{\rm apec}$ & $10^{-3}$ &  & $1.13^{+ 0.10}_{- 0.09}$ &  &  &  \\
$N_{\rm H}$ & $10^{24}$ cm$^{-2}$ &  & $10.00^{\rm peg}_{- 4.08}$ &  &  &  \\
$\Gamma$ &  &  & $2.00^f$ &  &  &  \\
$\theta_{\rm Incl}$ & degrees &  & $80^{\it f}$ &  &  &  \\
$A_{\rm S}$ & $10^{-2}$ &  & $6.31^{\rm peg}_{- 0.39}$ &  &  &  \\
$F_{\rm 0.5-10~keV}$ & $10^{-13}$ erg cm$^{-2}$ s$^{-1}$ &  & $8.85^{+ 0.15}_{- 0.16}$ &  &  &  \\
$\chi^2_{\rm r}$ (dof) &  &  & $1.15~(200)$ &  &  &  \\
\hline
\end{tabular}
\label{tab:XMMFits}
\begin{tablenotes}
\normalsize
\item Notes - 2010 epoch observations are from {\em Chandra} ACIS-S while remaining are from {\em XMM-Newton} EPIC pn.
$N_{\rm H}^{\rm gal}$ and $N_{\rm H}$ represent the Galactic and line-of-sight column density at the source redshift, respectively.
$\Gamma$ is photon index of power law emission. $kT$ denotes plasma temperature. $\tau_{\rm u}$ represents the upper limit on the ionisation timescale in {\scshape vpshock} model. `R'  and `F' denote reverse and forward shock components, respectively.
$A_{\rm S}$ is the normalisation of scattered component in {\scshape MYTorus} model. $^{f}$ Fixed value for a parameter.
\end{tablenotes}
\end{threeparttable}
\end{table*}
\subsection{CGX1}
\label{sec:CGX1}
CGX1 is an X-ray binary classified as an ULX with $L_{\rm 0.3-8.0~keV}$ in 
the range of 4.0 $\times$ 10$^{39}$ erg~s$^{-1}$ to 3.0 $\times$ 10$^{40}$~erg~s$^{-1}$ 
\citep{Bauer01,Esposito15,Qiu19}. 
The {\em Chandra} ACIS and {\em XMM-Newton} pn/MOS images detect CGX1 at 
RA = 14$^{h}$ 13$^{m}$ 12.21$^{s}$, DEC = -65$^{\circ}$ 20$^{\prime}$ 13$^{\prime\prime}$.7 (J2000), which is
15$^{\prime\prime}$ away from the AGN toward the northeast. 
The {\em Chandra} ACIS observations of higher angular resolution (0$^{\prime\prime}$.5) 
offer a clean detection of CGX1 with no contamination from the AGN and the diffuse emission.   
We extracted the {\em Chandra} ACIS spectrum of CGX1 using a circular aperture of 2$^{\prime\prime}$.4 radius.
Unlike {\em Chandra}, {\em XMM-Newton} pn spectra of CGX1 are extracted using a larger 
circular region of 7$^{\prime\prime}$.5 radius due to its larger PSF of 6$^{\prime\prime}$.0. 
Also, background determination in the {\em XMM-Newton} pn images is a little 
tricky due to the contamination from the AGN and the diffuse X-ray emission present around it.
Following \cite{Arevalo14}, we extracted the background spectra for CGX1 using an annular 
region around it, which excludes masked circular regions around the neighbouring CGX2 and AGN. 
The inner and outer radii of the annular region were set as 7$^{\prime\prime}$.5 and 30$^{\prime\prime}$, respectively. 
\par
\begin{figure*}
\centering
\includegraphics[angle=-90, width=\columnwidth, trim={0.0cm 0.0cm 0.0cm 0.0cm}, clip]{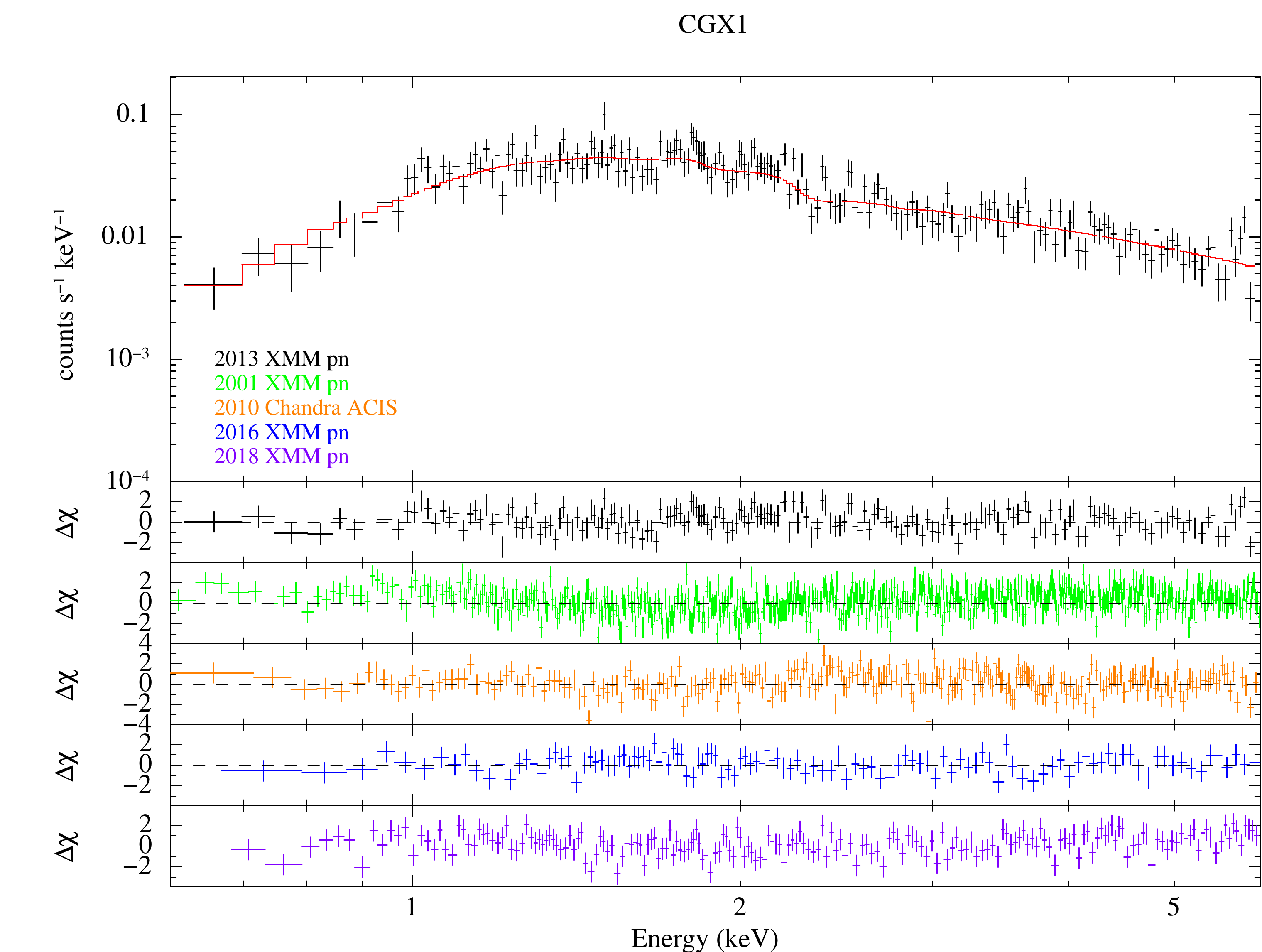}
\includegraphics[angle=-90, width=\columnwidth, trim={0.0cm 0.0cm 0.0cm 0.0cm}, clip]{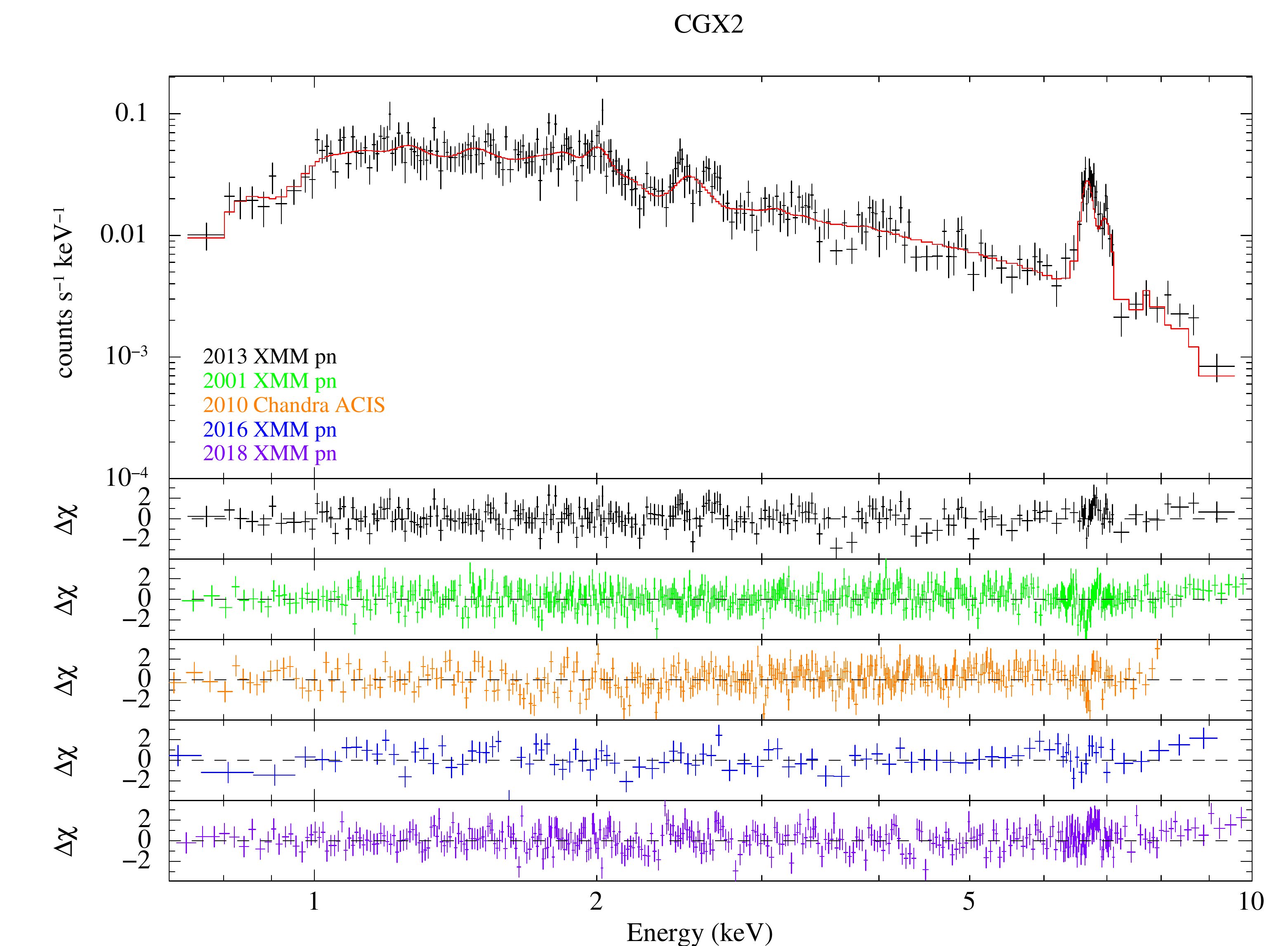}
\caption{{\it Left panel} : The 0.5$-$6.0 keV {\em XMM-Newton} pn 2013 spectrum and residuals of all epochs for CGX1. 
All epochs spectra are fitted with a simple absorbed power law. 
{\it Right panel} : The 0.5$-$10.0 keV {\em XMM-Newton} pn 2013 spectrum and residuals of all epochs for CGX2. 
All epochs spectra are fitted with a model characterised by the thermal emission from shock heated material. 
All epochs spectra are fitted with the same model.}
\label{fig:SpecContSrcs}
\end{figure*}

The {\em XMM-Newton} pn spectra begin to show rising background contamination above 6.0 keV 
\citep[see][]{Qiu19}. Therefore, we fit {\em XMM-Newton} pn spectra only in the 0.5$-$6.0 keV energy band. 
However, due to its higher angular resolution (0$^{\prime\prime}$.5) {\em Chandra} ACIS observations are  
not contaminated by the diffuse emission, and we consider the full 0.5$-$10 keV energy band, while fitting the
{\em Chandra} spectrum.  
We find that the 0.5$-$10 keV {\em Chandra} spectrum can be best$-$fitted with a simple absorbed power law model
with photon index ($\Gamma$) of 1.8. 
The absorption consists of two components, {\ie} galactic column density fixed to 5.6 $\times$ 
10$^{21}$ cm$^{-2}$ and intrinsic absorption due to the interstellar material (ISM) of the host galaxy.
Notably, the {\em XMM-Newton} pn spectra for all four epochs can also be fitted with an absorbed power law
(see Table~\ref{tab:XMMFits}).  
To attain a baseline model, we fixed the photon index to 1.8 while fitting the {\em XMM-Newton} spectra of different epochs. 
The absorbing column density and normalisations were left to vary. 
We find that absorbing column density is similar, in the range of $0.32^{+0.04}_{-0.04}~\times~ 10^{22}$ cm$^{-2}$ to
$0.42^{+0.06}_{-0.06}~\times~ 10^{22}$ cm$^{-2}$, during 2013, 2016 and 2018, while it is relatively high
(0.69$^{+0.03}_{-0.03}$ $\times$ 10$^{22}$ cm$^{-2}$) in 2001 epoch.  
Our spectral fits of CGX1 are consistent with the fact that the change in flux is 
due to variable absorption rather than the intrinsic change in the source itself  \citep{Qiu19}.   
We point out that the fitting of the 2001 spectrum can be improved ({\ie} the reduce ${\chi}^{2}$ changes from 1.28 to
1.11) by adding an ionised absorber ($\tt absori$).
However, the addition of the ionised absorber gives no significant improvement in the statistical fit of other epochs.
Therefore, to maintain a simple common baseline model across all the epochs,
we prefer to use only neutral absorbing material.
Figure~\ref{fig:SpecContSrcs} shows the fitted spectra and residuals for all five different epochs.
In Table~\ref{tab:XMMFits}, we list the best$-$fitted parameters for different epochs.
\begin{figure}
\centering
\includegraphics[angle=-90, width=\columnwidth, trim={0.0cm 0.0cm 0.0cm 0.0cm}, clip]{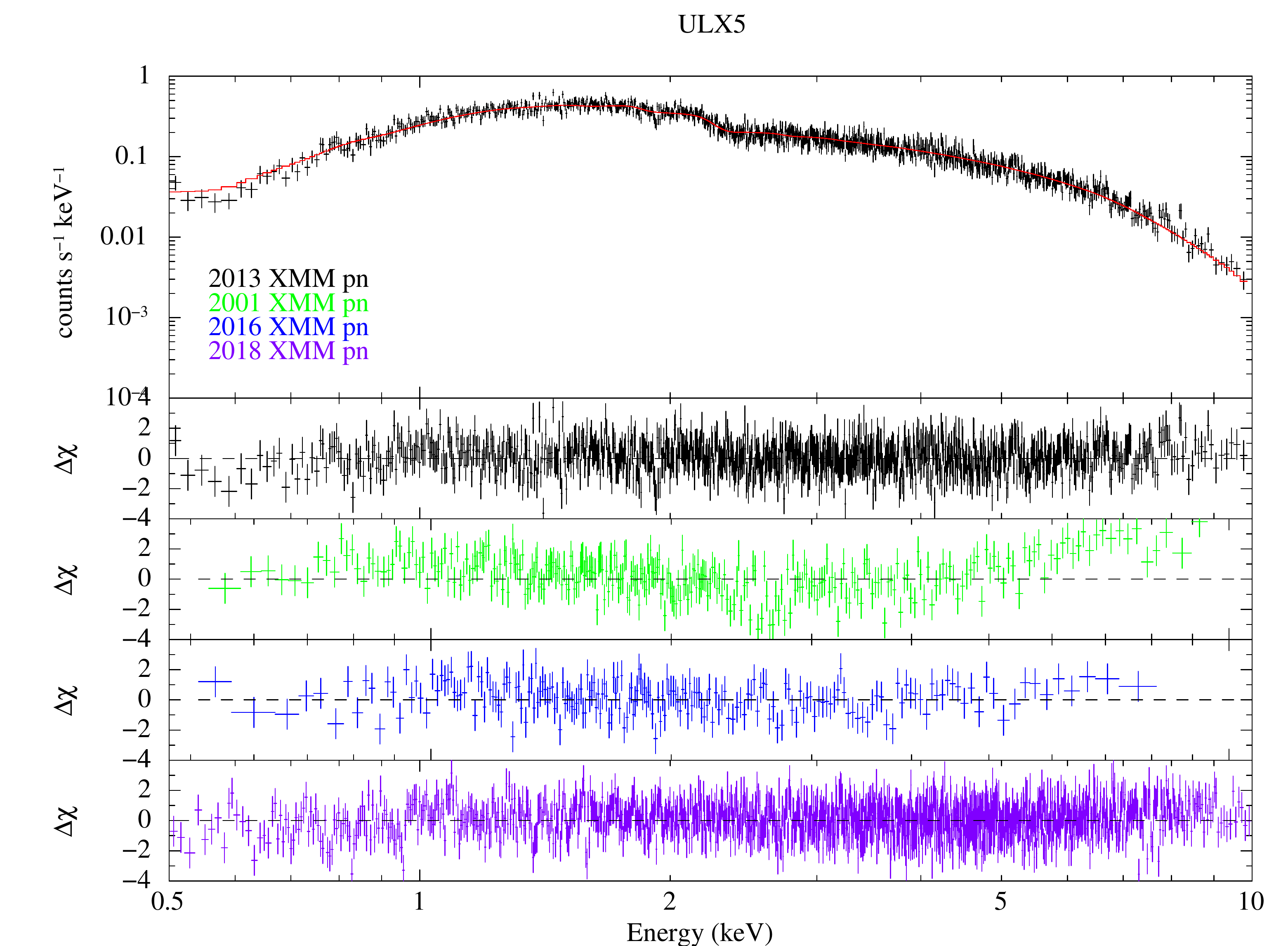}
\caption{The 0.5$-$10 keV {\em XMM-Newton} pn 2013 spectrum and residuals for all four epochs for ULX5. 
All epochs spectra are fitted with {\scshape diskbb} model.}
\label{fig:SpecULX5}
\end{figure}
\begin{figure}
\centering
\includegraphics[angle=-90, width=\columnwidth, trim={0.0cm 0.0cm 0.0cm 0.0cm}, clip]{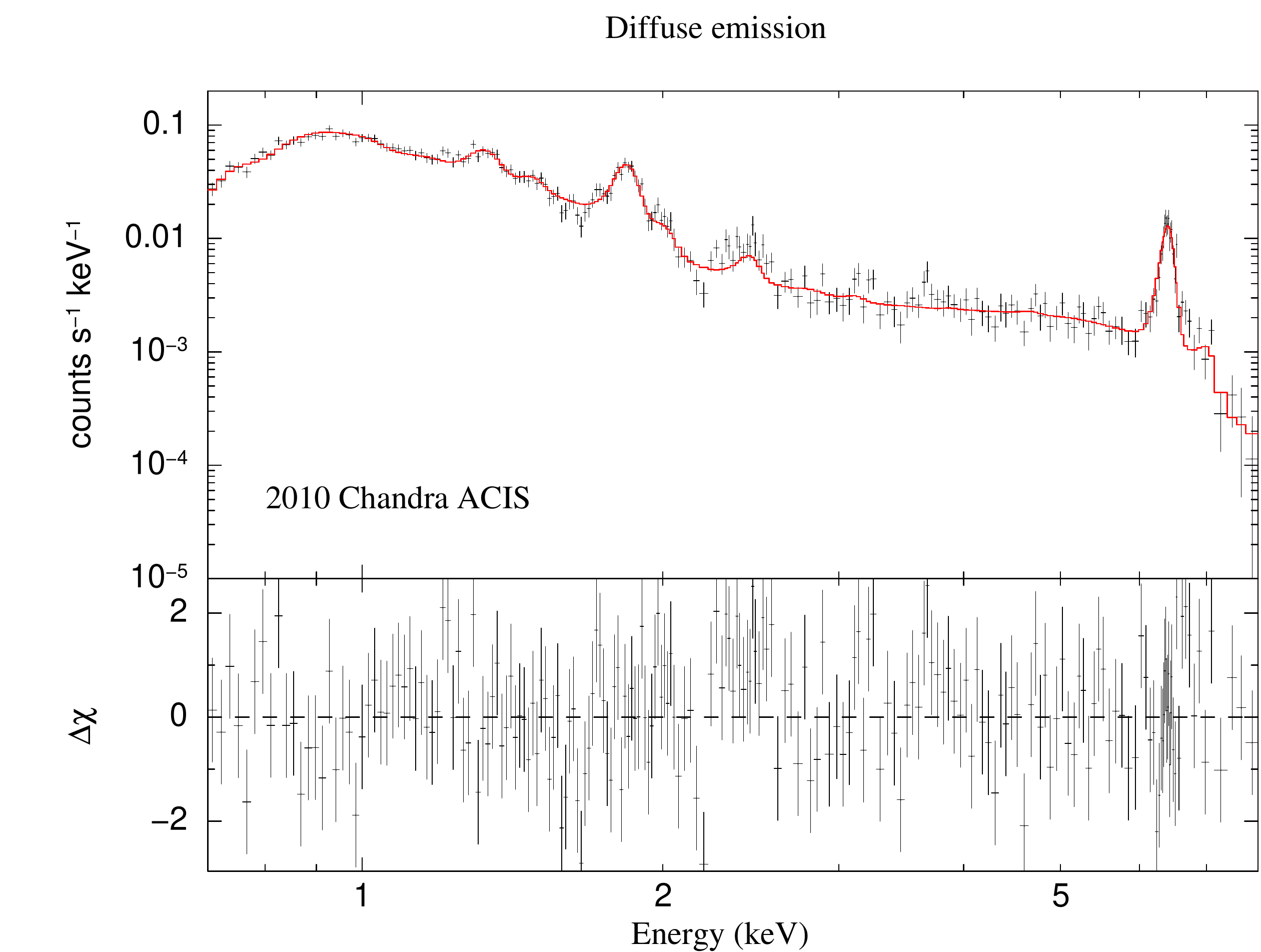}
\caption{The 0.7-8.0 keV {\em Chandra} spectrum of diffuse X-ray emission fitted with a model consisting 
of a soft component represented by {\scshape apec} plus a hard component represented by Compton reflection and emission lines.}
\label{fig:SpecDiff}
\end{figure}
\subsection{CGX2}
\label{sec:CGX2}
CGX2 is known to be a young supernova remnant and the X-ray emission is interpreted as the
thermal X-ray emission arising from a shock-heated plasma produced by the interaction of outflowing supernova 
ejecta with the circum-stellar material (CSM) \citep{Quirola19}. 
The CGX2 is located at RA = 14$^{\rm h}$ 13$^{\rm m}$ 10$^{\rm s}$.01, DEC = -65$^{\circ}$ 20$^{\prime}$ 44$^{\prime\prime}$.4, 25$^{\prime\prime}$ south of AGN. 
In {\em Chandra} ACIS image of 2010, CGX2 is neatly separated from the diffuse emission around AGN. 
We extracted the source spectrum by considering a circular extraction
region of 2$^{\prime\prime}$.4 radius centred at CGX2. 
The background spectrum was extracted from an annular region centered at CGX2 with inner and outer radii 
of 5$^{\prime\prime}$.0 and 7$^{\prime\prime}$.0, respectively. 
The {\em XMM-Newton} pn spectra of different epochs 
were extracted using a circular extraction region of 10$^{\prime\prime}$ radius. 
The background spectra were extracted in a manner 
similar to that for CGX1, {\ie} an annulus centred at CGX2 with the exclusion of masked regions around CGX1 and AGN.     
\par 
We find that the 0.5$-$10 keV {\em Chandra} ACIS and {\em XMM-Newton} pn/MOS spectra can be best$-$fitted
with a combination of two {\scshape vpshock} models. 
The {\scshape vpshock} model characterises X-ray emission arising from a constant temperature 
shock-heated plasma \citep{Borkowski01}. 
The best fit of the {\em XMM-Newton} pn spectrum for the 2013 epoch is obtained with a model consisting of 
two {\scshape vpshock} components with plasma temperatures ($kT$) of $2.98^{+1.81}_{-0.87}$ keV and $7.89^{+0.75}_{-0.69}$ keV (see Table~\ref{tab:XMMFits}). The two plasma components can be interpreted as
the forward and reverse shock emission. The two plasma emission components are absorbed with 
a column density ($N_{\rm H}$) in the range of $0.23^{+0.08}_{-0.08} - 0.70^{+0.08}_{-0.08}~\times~10^{22}$~cm$^{-2}$
in addition to the galactic column density of 5.6 $\times$ 10$^{21}$~cm$^{-2}$. 
For simplicity, we assumed that 
both plasma components suffer from the same amount of absorption, and hence $N_{\rm H}$ of both components 
were tied together. We find that the {\em XMM-Newton} pn and {\em Chandra} spectra of other
epochs can be fitted reasonably well with the same model 
(see Table~\ref{tab:XMMFits}). 
To maintain the same baseline model with minimum free parameters, we fixed plasma temperatures to 3.0 
keV and 8.0 keV, values similar to that obtained from the 2013 epoch {\em XMM-Newton} pn spectrum. 
We note that, in addition to the plasma temperature, the {\scshape vpshock} model includes 
ionisation timescale $\tau$ = $n_{e}t$, where $n_{e}$ is the electron density and $t$ is the time since the plasma 
was shocked, individual atomic abundances for various elements such as  He, C, N, O, Ne, Mg, Si, S, Ar, Ca, Fe, Ni. 
The {\em Chandra} and {\em XMM-Newton} spectra show the presence of various emission lines, including strong 
Fe lines. We obtain improved spectral fits by adding multiple narrow Gaussian lines that account for
broad asymmetric profiles of emission lines present at various energies in the range of 0.8 to 7.8 keV.
In our best fits, most of the emission lines except Fe lines can be accounted for by using abundances fixed to Solar values.
As expected, our fitted parameters are broadly consistent with those reported by \cite{Quirola19}, who 
characterised the continuum emission with the same model but convolved it with the {\scshape shellblur} model. 
The {\scshape shellblur} model accounts for asymmetry in emission line profiles, clearly seen in the {\em Chandra} grating spectra.    
\subsection{ULX5}
\label{sec:ULX5}
The ULX5 
located at RA = 14$^{\rm h}$ 12$^{\rm m}$ 39$^{s}$, DEC = -65$^{\circ}$ 23$^{\prime}$ 34$^{\prime\prime}$ 
at the distance of 4$^{\prime}$.5 away from AGN is clearly detected in the 
{\em XMM-Newton} pn and {\em NuSTAR} images. 
ULX5 lying in the outskirts of Circinus galaxy is known to be a variable source 
with X-ray luminosity in the range of 4.5 $\times$ 10$^{39}$ to 1.8 $\times$ 10$^{40}$ erg s$^{-1}$
\citep[see][]{Walton13,Mondal21}.
We extracted {\em XMM-Newton} pn spectra using a circular aperture of 30$^{\prime\prime}$ radius.
The background spectra were extracted using a source-free region in the same chip. 
We find that 0.5$-$10 keV multi-epoch {\em XMM-Newton} pn/MOS spectra can be best$-$fitted
with the {\scshape diskbb} model characterising 
emission from an accretion disk assumed to be consisting of multiple blackbody components. 
The spectral modelling of different epochs yields the temperature of the accretion disk in the range of 1.0 to 2.0 keV 
(see Table~\ref{tab:XMMFits}). 
The 0.5$-$10~keV flux and accreting disc temperature suggest ULX5 to be in a high state in the 2013 and 2018 epochs, while it appears to be in a low state during the 2001 and 2016 epochs.
Figure~\ref{fig:SpecULX5} shows multi-epoch spectral fits and residuals.  
\subsection{Diffuse soft X-ray emission}
\label{sec:Diffuse}
The {\em Chandra} ACIS image shows extended diffuse X-ray emission around the AGN with an apparent
elongation towards the north$-$west direction (see Figure~\ref{fig:ChandraIm}).
We extracted the spectrum using a polygon region such that it includes most of the extended emission and excludes AGN
and X-ray point sources. 
We find that the  0.7$-$8.0 keV {\em Chandra} X-ray spectrum can be fitted with a model consisting of a 
soft component represented by the {\scshape apec} model characterising the emission from hot gas, and a hard  
component representing the scattered emission from AGN (see Figure~\ref{fig:SpecDiff}). 
The hard component can be described by the scattered power law component in the {\scshape MYTorus} model.
Our best-fitted model gives a plasma temperature of {$0.82^{+0.02}_{-0.02}$} keV and a steep power law ($\Gamma$ = 2.0) emission
Compton-scattered from a nearly edge-on torus (${\theta}_{\rm incl}$ = {$80^{\circ}$}) with
the line-of-sight column density of 10$^{25}$ cm$^{-2}$, which allows no directly transmitted component 
in the {\em Chandra} energy band (see Table~\ref{tab:XMMFits}). 
We obtain the best fit (reduced ${\chi}^{2}$ = 1.15) by keeping the elements abundance in {\scshape apec} model as a 
free parameter and by adding a few unresolved Gaussian emission lines at 0.74 keV, 1.3 keV and 1.8 keV. 
The elements abundance is only 0.2 times of the Solar value.
\par
Our {\em Chandra} X-ray spectrum is similar to that reported in \cite{Arevalo14}, who 
analysed 0.7$-$8.0 keV {\em Chandra} spectra of different regions, {\ie} ionisation cone, circumnuclear region, 
central annulus, large-scale extended emission, and found similar spectra for all the regions. 
\cite{Mingo12} highlighted the morphological correspondence between radio lobes and diffuse extended 
X-ray emission at kpc-scales and favoured a 
scenario in which extended diffuse X-ray emission is mainly arising from the shock-heated gas produced 
via the interaction of the radio jet with the surrounding gas. They found that the soft X-ray spectrum 
can be fitted well with the {\scshape apec} model yielding gas temperature ($kT$) of 0.74 keV and a low 
elements abundance of 0.15 $Z_{\odot}$. 
We note that our spectral parameters are consistent with the previous studies 
\citep[{e.g.,}][]{Mingo12,Arevalo14}.  
\subsection{Contamination model}
\label{sec:ContModel}
To account for the contributions from all contaminating sources while performing broadband spectral fittings, we formulate a contamination model. 
We define contamination model as the sum of the models characterising contaminants (CGX1, CGX2 and extended X-ray emission), 
{\ie} \textsc{tbabs $\times$ (tbabs $\times$ po + tbabs $\times$ vpshock + tbabs $\times$ vpshock + apec + lines)}. 
We note that the hard component of diffuse extended X-ray emission is a Compton-scattered AGN component which 
would be included in the physically motivated models considering the scattered emission from AGN. 
Hence, we do not add a scattered power law component to our contamination model. 
In the case of {\em BeppoSAX}, {\em Suzaku} and {\em AstroSat} spectral modelling, we include the contribution from ULX5,
and hence, the contamination model is defined as  \textsc{ tbabs $\times$ (tbabs $\times$ po + tbabs $\times$ vpshock + tbabs $\times$ vpshock + apec + lines + tbabs $\times$ diskbb)}. 
We note that, for epochs with no simultaneous {\em XMM-Newton} observations, we consider the 
fact that the multi-epoch spectra of a contaminant can be fitted with the same model and variability seen 
across different epochs can be accounted for by a varying normalisation (see Table~\ref{tab:XMMFits}).  
Thus, while modelling broadband spectra, we fixed spectral shapes and parameters of contaminants 
but varied their normalisations within a range observed from their multi-epoch spectral fittings. 
The contribution from extended diffuse X-ray emission is likely to remain constant. 
Hence, we keep all the spectral parameters of the soft component of the diffuse emission fixed.
\section{Broadband X-ray spectral modelling}
\label{sec:BBmodels}
We modelled the broadband X-ray spectra of Circinus using all ten epochs of hard X-ray observations
(see Table~\ref{tab:ObsLog}). We avoid soft X-ray data points below 3.0 keV to mitigate the
effects of off-nuclear contaminating sources. 
The exclusion of X-ray data below 3.0 keV also brings uniformity 
in terms of spectral coverage for all the multi-epoch spectra, considering that only {\em NuSTAR} spectra 
covering the 3.0$-$79 keV energy band are available for four epochs (see Table~\ref{tab:ObsLog}).
Thus, broadband spectra are limited to 3.0$-$79 keV whenever {\em NuSTAR} observations are used, even
when the simultaneous {\em XMM-Newton} observations are available.
The {\em BeppoSAX}, {\em Suzaku} and {\em AstroSat} spectra cover nearly 3.0$-$100 keV energy range.
\par
We fitted the spectra of all ten epochs together, considering that the joint fit is useful for reducing the uncertainties in the spectral parameters and for breaking degeneracies between different parameters \citep{Balokovic21,Saha22}.
The joint fit of multi-epoch spectra can be considered equivalent to the spectrum integrated over
a long period of time. Therefore, we can easily identify spectral parameters that remain constant over
a long timescale. A parameter variable across different epochs can also be identified by allowing
it to vary across epochs. Therefore, with the multi-epoch joint spectral modelling we aim to place better constraints on geometrical
parameters such as torus average column density, covering factor and inclination angle.
We also aim to probe the viability of variable line-of-sight column
density ($N_{\rm H,LOS}$).
To minimise the model-dependent effects, we
explore three different physically-motivated models. The details of the spectral fittings with these
models are given below.
\subsection{Spectral modelling with {\scshape MYTORUS} model}
\label{sec:MYTorus}
The {\scshape MYTorus} model considers reprocessing material distributed in a toroidal geometry with 
a circular cross-section and uniform density \citep{Murphy09}. 
The opening angle of the torus (${\theta}_{\rm tor}$) is fixed to $60^{\circ}$,  
which gives a covering factor of (cos(${\theta}_{\rm tor})$) $0.5$.
The inclination angle of the torus (${\theta}_{\rm incl}$), {\ie} the angle between the torus symmetric 
axis and the line-of-sight, is a free parameter, and it can have any value in the range of 
0$^{\circ}$ (a face-on torus) to 90$^{\circ}$ (an edge-on torus). 
The {\scshape MYTorus} model considers different spectral components, {\ie} AGN intrinsic emission 
transmitted through obscuring material, the scattered emission from the reprocessing material 
toward the line-of-sight, and the fluorescent emission lines, in a self-consistent manner. 
The three components are denoted as MYTZ (transmitted component), 
MYTS (Compton-scattered continuum), and MYTL (fluorescent emission lines). 
The MYTZ, an energy-dependent multiplicative factor applied to the intrinsic continuum, accounts for 
the line-of-sight obscuration. In principle, the intrinsic continuum can have any spectral shape. 
We assumed a power law spectral shape which is consistent 
with the inherent assumption of the {\scshape MYTorus} model that the scattered continuum and ﬂuorescent 
emission-line components are reprocessed emission of a power law continuum. 
The scattered component can have cutoff energy in the range of 100~keV to 500~keV.  
The MYTL component fits ﬂuorescent emission lines Fe K$\alpha$ and Fe K$\beta$ 
at 6.4 keV and 7.06 keV, respectively. 
\par
To fit the joint spectra of different epochs, we begin with the {\scshape MYTorus} model plus
the contamination model. The {\scshape MYTorus} model is aimed at fitting the AGN spectral components, while the contamination model accounts for the contribution from the contaminant sources (see Section~\ref{sec:ContModel}).
We considered the transmitted component (MYTZ) and 
the Compton-scattered component (MYTS). In XSPEC notation, the model can be expressed as \textsc{$c_{1}$ $\times$ { tbabs (MYTZ $\times$ cutoffpl}~+~$c_{2}$ $\times$ {MYTS}} + lines + contamination~model).
First, we used the {\scshape MYTorus} model in the coupled configuration by tying together various parameters 
({\ie} photon index, column density and normalisations) of all three components. 
This configuration assumes that the scattered and the emission line components result from the
reprocessing of the intrinsic power law emission from a uniform torus.
While performing the fit, we kept photon index ($\Gamma$),
column density ($N_{\rm H}$), inclination angle (${\theta}_{\rm incl}$) and normalisations as free parameters. 
We note that the {\scshape MYTorus} model introduces a cutoff on the scattered component by using 
different tables with termination energies fixed to 100 keV, 160 keV, 200 keV, 300 keV, 400 keV and 500 keV.
To have consistency between the scattered and the transmitted components, we used a cutoff power law for the
transmitted component with cutoff energy ($E_{\rm cut}$) fixed to the termination energy of 
the scattered component. 
Also, we notice that the {\em NuSTAR} spectra show Ni absorption edge 
at 8.3 keV, which is not included in the {\scshape MYTorus} scattered component. 
Therefore, while fitting the {\em NuSTAR} spectra, we added Ni edge using the {\scshape zedge} model and kept its 
optical depth and normalisation as free parameters. 
The cross-normalisation factors between different instruments are accounted by a multiplicative factor to
the cumulative model and are allowed to vary for all instruments with respect to the {\em XMM-Newton} data of the
2013 February 03 epoch, which is fixed to one.
In Table~\ref{tab:MYToFits}, we list the cross-normalisation factors for different instruments and find
them to be consistent with the values reported in the literature \citep[{e.g.,}][]{Madsen17}.
\par
To fit the multi-epoch spectra, we first assumed no variability across epochs for any of the parameters.
So, all the spectral parameters of different epochs were tied together with an assumption of no intra-epoch variability in all the parameters (see case~(i) in Table~\ref{tab:MYToFits}).
We find that the {\scshape MYTorus} model reproduces Compton hump peaking at 30~keV,
but it leaves significant residuals in the soft band.
The 6.4 keV emission line also shows residuals for the {\em XMM-Newton} data points,
which can be understood due to the fact that the spectral resolution of the {\em XMM-Newton} (150 eV)
is nearly 2.5 times better than that for {\em NuSTAR}, which has 400 eV spectral resolution at 6.4 keV.
The addition of an unresolved emission line only to the {\em XMM-Newton} data provides an improved fit.
In the case of {\em BeppoSAX}, {\em Suzaku}, and {\em AstroSat} spectral fittings, we added unresolved emission
lines to the MECS, XISs and SXT spectra.
The addition of unresolved emission lines at various energies ranging from 3.0 keV to 7.45 keV accounts for
the residuals seen in the soft band.
In general, energies and normalisations of these emission lines are consistent with the 
previous studies \citep{Sambruna01,Massaro06,Arevalo14}, which identify them as Ar, Ca, Cr, Fe 
and Ni lines of different ionisations, with several of them being He-like and H-like ions.
Except for Fe lines, all other emission lines are less prominent with the equivalent widths 
in the range of 20 eV to 150 eV and normalisations of the order of 10$^{-4}$ to 10$^{-6}$. 
\par 
\begin{table*}
\centering
\begin{adjustbox}{width=\textwidth}
\renewcommand{\arraystretch}{1.25}
\begin{threeparttable}
\caption{The joint fit spectral parameters using {\scshape MYTorus} model}
\begin{tabular}{ccccccccccc}
\hline
Parameters & 1998-03-13 & 2001-01-07 & 2006-07-21 & 2013-01-25 & 2013-02-02 & 2013-02-03 & 2013-02-05 & 2016-08-23  & 2020-01-28 & 2020-11-26 \\
    & ($Beppo$) & ($Beppo$) & ($Suzaku$) & ($Nu$)   &   ($Nu$)   & ($X+Nu$)    & ($Nu$) & ($X+Nu$)  &
    ({${AstroSat}$}) & ($Nu$)     \\
    &    (1)     &    (2)    &    (3)    &  (4)   &  (5)   & (6 ) & (7)  & (8)   &  (9)   &  (10)   \\
\hline
\multicolumn{11}{c}{~Model: {\scshape MYTorus} + lines + contamination}  \\
\hline
\multicolumn{11}{c}{\underline{case~(i) : all parameters varied but tied across epochs}}  \\
$\Gamma$                   & \multicolumn{10}{c}{{$2.28^{+0.01}_{-0.02}$}}  \\
$N_{\rm H,LOS}$           & \multicolumn{10}{c}{{$6.10^{+0.20}_{-0.17}$}}  \\
${\theta}_{\rm incl}$      & \multicolumn{10}{c}{{$76.9^{+0.5}_{-0.5}$}} \\
$A_{\rm Z}~(10^{-9})$  & \multicolumn{10}{c}{{$1.32^{+0.11}_{-0.12}$}} \\
$A_{\rm S}$  & \multicolumn{10}{c}{{$1.22^{+0.10}_{-0.11}$}} \\
$E_{\rm cut}$              & \multicolumn{10}{c}{{$160^f$}} \\
${\chi}_{r}$(dof)          & \multicolumn{10}{c}{{$1.123~(8107)$}} \\
$C_{\rm XMM}$  &         &       &         &      &     &  {$1.0^{f}$} &      & {$0.96^{+0.01}_{-0.01}$} &     &     \\
$C_{\rm NuA}$  &         &       &   & {$1.30^{+0.01}_{-0.01}$} & {$1.23^{+0.02}_{-0.02}$} & {$1.23^{+0.01}_{-0.01}$} & {$1.19^{+0.01}_{-0.01}$} &  {$1.34^{+0.02}_{-0.02}$}   &     & {$1.20^{+0.01}_{-0.01}$} \\
$C_{\rm NuB}$   &          &       &     &  {$1.35^{+0.01}_{-0.01}$} & {$1.27^{+0.02}_{-0.02}$} & {$1.26^{+0.01}_{-0.01}$} & {$1.22^{+0.01}_{-0.01}$} & {$1.40^{+0.02}_{-0.02}$} &      & {$1.21^{+0.01}_{-0.01}$}    \\
$C_{\rm XIS0}$ & {$0.81^{+0.01}_{-0.01}$} (ME) & {$0.77^{+0.01}_{-0.01}$} (ME) & {$0.94^{+0.01}_{-0.01}$} &         &     &   &      &     & {$0.72^{+0.04}_{-0.04}$} (S) &     \\
$C_{\rm XIS1}$ &           &       & {$0.96^{+0.01}_{-0.01}$} &         &     &   &     &     &     &     \\
$C_{\rm XIS2}$ &           &       & {$0.91^{+0.01}_{-0.01}$} &         &     &   &     &     &     &     \\
$C_{\rm XIS3}$ &           &       & {$0.91^{+0.01}_{-0.01}$} &         &     &   &     &     &     &     \\
$C_{\rm PIN}$  & {$1.18^{+0.03}_{-0.03}$} (P) & {$1.03^{+0.03}_{-0.03}$} (P) & {$1.45^{+0.02}_{-0.02}$} &         &     &   &      &     &  {$0.91^{+0.05}_{-0.05}$} (LX)   &     \\
$C_{\rm GSO}$  &           &       & {$1.56^{+0.26}_{-0.26}$} &         &     &   &    &     & {$1.52^{+0.25}_{-0.25}$} (CZ) &     \\
\hline
\multicolumn{11}{c}{\underline{case~(ii) : only normalisation untied across epochs}}  \\
$\Gamma$                   & \multicolumn{10}{c}{{$2.27^{+0.02}_{-0.02}$}}  \\
$N_{\rm H,LOS}$ & \multicolumn{10}{c}{{$6.03^{+0.18}_{-0.18}$}}  \\
${\theta}_{\rm incl}$      & \multicolumn{10}{c}{{$76.7^{+0.5}_{-0.5}$}} \\
$A_{\rm Z}~(10^{-9})$  & {$1.32^{+0.18}_{-0.16}$} & {$1.22^{+0.22}_{-0.20}$} & {$1.32^{+0.12}_{-0.12}$} & {$1.25^{+0.13}_{-0.12}$} & {$1.27^{+0.14}_{-0.13}$} & {$1.28^{+0.12}_{-0.11}$} & {$1.29^{+0.14}_{-0.12}$} &  {$1.36^{+0.14}_{-0.13}$} & {$1.45^{+0.42}_{-0.31}$} & {$1.29^{+0.13}_{-0.12}$} \\
$A_{\rm S}$ & {$1.23^{+0.17}_{-0.15}$} & {$1.05^{+0.19}_{-0.17}$} & {$1.23^{+0.12}_{-0.11}$} & {$1.10^{+0.11}_{-0.10}$} & {$1.14^{+0.12}_{-0.11}$} & {$1.16^{+0.11}_{-0.10}$} & {$1.17^{+0.12}_{-0.11}$} &  {$1.30^{+0.13}_{-0.12}$} & {$1.48^{+0.43}_{-0.32}$} & {$1.17^{+0.12}_{-0.11}$} \\
$E_{\rm cut}$              & \multicolumn{10}{c}{{$160^f$}} \\
${\chi}_{r}$(dof)          & \multicolumn{10}{c}{{$1.116~(8089)$}} \\
\hline
\multicolumn{11}{c}{\underline{case~(iii) : only $N_{\rm H,LOS}$ untied across epochs}}  \\
$\Gamma$                   & \multicolumn{10}{c}{{$2.28^{+0.01}_{-0.01}$}}  \\
$N_{\rm H,LOS}$ & {$6.00^{+1.35}_{-1.34}$} & {$6.28^{+1.09}_{-1.17}$} & {$4.43^{+0.22}_{-0.32}$} & {$6.47^{+0.20}_{-0.19}$} & {$6.33^{+0.25}_{-0.20}$} & {$6.39^{+0.21}_{-0.20}$}    & {$6.09^{+0.20}_{-0.20}$} & {$5.28^{+0.47}_{-0.18}$} &
{$4.86^{+1.15}_{-0.84}$} & {$6.14^{+0.47}_{-0.16}$} \\
${\theta}_{\rm incl}$      & \multicolumn{10}{c}{{$77.2^{+0.8}_{-0.5}$}} \\
$A_{\rm Z}~(10^{-9})$  & \multicolumn{10}{c}{{$1.36^{+0.20}_{-0.11}$}} \\
$A_{\rm S}$ & \multicolumn{10}{c}{{$1.29^{+0.20}_{-0.10}$}} \\
$E_{\rm cut}$              & \multicolumn{10}{c}{{$160^f$}} \\
${\chi}_{r}$(dof)          & \multicolumn{10}{c}{{$1.101~(8098)$}} \\
\hline
\multicolumn{11}{c}{\underline{case~(iv) : $N_{\rm H,LOS}$ and normalisations untied across epochs}}  \\
$\Gamma$                   & \multicolumn{10}{c}{{$2.28^{+0.01}_{-0.01}$}}  \\
$N_{\rm H,LOS}$ & {$6.17^{+1.36}_{-1.43}$} & {$6.23^{+1.17}_{-1.22}$} & {$4.48^{+0.22}_{-0.34}$} & {$6.43^{+0.21}_{-0.32}$} & {$6.37^{+0.32}_{-0.32}$} & {$6.39^{+0.22}_{-0.33}$}    & {$5.97^{+0.28}_{-0.36}$} & {$5.14^{+0.38}_{-0.22}$} &
{$5.32^{+2.80}_{-1.84}$} & {$6.18^{+0.19}_{-0.40}$} \\
${\theta}_{\rm incl}$      & \multicolumn{10}{c}{{$77.1^{+0.6}_{-0.7}$}} \\
$A_{\rm Z}~(10^{-9})$  & {$1.37^{+0.13}_{-0.12}$} & {$1.27^{+0.20}_{-0.18}$} & {$1.35^{+0.03}_{-0.03}$} & {$1.33^{+0.04}_{-0.04}$} & {$1.35^{+0.08}_{-0.08}$} & {$1.34^{+0.03}_{-0.03}$} & {$1.32^{+0.07}_{-0.07}$} &  {$1.32^{+0.07}_{-0.05}$} & {$1.43^{+0.62}_{-0.52}$} & {$1.35^{+0.04}_{-0.04}$} \\
$A_{\rm S}$ & {$1.31^{+0.18}_{-0.18}$} & {$1.14^{+0.16}_{-0.20}$} & {$1.28^{+0.12}_{-0.15}$} & {$1.25^{+0.14}_{-0.14}$} & {$1.28^{+0.16}_{-0.16}$} & {$1.27^{+0.13}_{-0.13}$} & {$1.23^{+0.14}_{-0.15}$} &  {$1.22^{+0.18}_{-0.14}$} & {$1.44^{+0.67}_{-0.59}$} & {$1.28^{+0.14}_{-0.19}$} \\
$E_{\rm cut}$              & \multicolumn{10}{c}{{$160^f$}} \\
${\chi}_{r}$(dof)          & \multicolumn{10}{c}{{$1.102~(8080)$}} \\
\hline
\end{tabular}
\label{tab:MYToFits}
\begin{tablenotes}
\normalsize
\item Notes - $Beppo$ : {\em BeppoSAX}; $X+Nu$ : {\em XMM-Newton} plus {\em NuSTAR}; $Nu$ : {\em NuSTAR}.
$C_{\rm XMM}$, $C_{\rm NuA}$, $C_{\rm NuB}$, $C_{\rm XIS0}$, $C_{\rm XIS1}$, $C_{\rm XIS2}$, $C_{\rm XIS3}$, $C_{\rm PIN}$,
$C_{\rm GSO}$ represent cross-normalisation factors for {\em XMM-Newton}, {\em NuSTAR} FPMA, {\em NuSTAR} FPMB,
{\em Suzaku} XIS0, XIS1, XIS2, XIS3, PIN and GSO, respectively. The cross-normalisation factors for {\em BeppoSAX} MECS, PDS,
{\em AstroSat} SXT, LAXPC and CZTI are indicated with `ME', `P', `S', `LX' and `CZ', respectively.
The units of column densities, inclination angle (${\theta}_{\rm incl}$) and normalisations
are 10$^{24}$~cm$^{-2}$, degrees and erg~cm$^{2}$~s$^{-1}$~keV$^{-1}$, respectively.
$^{f}$ Fixed value of a parameter.
\end{tablenotes}
\end{threeparttable}
\end{adjustbox}
\end{table*}
We note that keeping all the parameters tied across epochs gives an acceptable fit with
{${\chi}^{2}_{r}$} (dof) = 1.123 (8107) (see case (i) in Table~\ref{tab:MYToFits}).
However, as expected, the fit statistics improve slightly with {${\chi}^{2}_{r}$} (dof) = 1.116 (8089) if normalisations are varied across epochs. The fit statistics show further improvement if $N_{\rm H,LOS}$
is considered as a varying parameter across epochs (see case (iii) in Table~\ref{tab:MYToFits}).
We note that, an equally good fit is obtained if normalisations across epochs are
tied but $N_{\rm H,LOS}$ is kept variable (see case (iv) in Table~\ref{tab:MYToFits}).
Therefore, our joint spectral fitting favours the variability in $N_{\rm H,LOS}$ across epochs.
It is worth noting that, the fit statistics show
no significant improvement if other parameters such as photon index ($\Gamma$), inclination angle (${\theta}_{\rm incl}$)
are kept variable across epochs.
Thus, our spectral modelling with the {\scshape MYTorus} model shows that, in all epochs, AGN emission can be characterised with a steep power law spectrum ({$\Gamma~=~2.28^{+0.01}_{-0.01}$})
piercing through a Compton-thick line-of-sight column density
({$N_{\rm H,LOS}$} in the range of {$4.48^{+0.22}_{-0.34}~\times~10^{24}$} cm$^{-2}$ to {$6.43^{+0.21}_{-0.32}~\times~10^{24}$} cm$^{-2}$) in a nearly edge-on ({${\theta}_{\rm incl}$} = 77$^{\circ}$) torus.
Further, we find that the transmitted component vanishes with its
normalisation nearly 10$^{9}$ times lower than that of the scattered component.
The absence of the transmitted component can be ascribed to a heavily Compton-thick line-of-sight
column density wherein hard X-ray spectrum ($>$ 10 keV) is fully accounted for by the scattered component.
We note that our results are consistent
with \cite{Arevalo14}, who modelled $2.0-79$ keV combined spectra of {\em XMM-Newton}, {\em NuSTAR} spectra
and {\em Swift}/BAT and reported a Compton scattered dominated steep power law ($\Gamma$ = 2.2 $-$ 2.4)
spectrum from an obscured AGN with torus equatorial column density of $N_{\rm H}$ = 6$-$10 $\times$ 10$^{24}$ cm$^{-2}$.
\par
To examine the possibility of a patchy torus in which transmitted and scattered components encounter 
different column densities, we performed spectral fitting by untying the column densities 
of the transmitted and the scattered components. We obtained nearly the same fit statistics with
nearly the same parameters, although, globally averaged column density
({$N_{\rm H,avg}~=~5.70^{+0.30}_{-0.30}~\times~10^{24}$}~cm$^{-2}~-~10^{25}$~cm$^{-2}$)
implied from the scattered component is significantly lower than the line-of-sight column 
density ($N_{\rm H,LOS}$ $\geq$ 10$^{25}$~cm$^{-2}$). In fact, $N_{\rm H,LOS}$ is encountering the  
upper limit allowed in the {\scshape MYTorus} model, and an actual value can be even higher. 
A significantly higher line-of-sight column density than the global average column density 
indicates a patchy torus around the AGN.
However, we point out that the line-of-sight column density, attaining the upper limit of its value allowed
by the model, remains unconstrained. To overcome this issue, we attempted to fit our multi-epoch spectra 
using other models described below. 
\subsection{Spectral modelling with {\scshape BORUS02} model}
\label{sec:borus}
The {\scshape borus02} model \citep{Balokovic18} assumes uniform-density reprocessing matter having a 
spherical geometry with two conical polar cut-outs. 
The {\scshape borus02} model consists of a Compton-scattered continuum and fluorescent emission lines. 
The intrinsic continuum is accounted separately by a cutoff power law multiplied by 
a line-of-sight absorbing column density ({\scshape tbabs}) and the 
Compton scattering losses ({\scshape cabs}) occurred along the line-of-sight.
Unlike the {\scshape MYTorus} model, this model considers 
covering factor ($f_{c}$) that can vary in the range of 0.1 to 1.0, corresponding to the torus 
opening angle ($\theta_{\rm tor}$) in $0^{\circ}$ to 84$^{\circ}$ range. 
The torus inclination angle ($\theta_{\rm incl}$) varies in the range of 0$^{\circ}$ to 90$^{\circ}$. 
The {\scshape borus02} model can account for a patchy environment by allowing a line-of-sight column 
density ($N_{\rm H,LOS}$) to be different than the average column density of the torus 
($N_{\rm H,tor}$). 
\par 
We fitted the joint spectra of all ten epochs using the {\scshape borus02} model that
can be expressed as 
\textsc{$c_{\rm 1}$ ${\times}$ {tbabs} ({ztbabs} {$\times$} {cabs} {$\times$} {cutoffpl} + {borus02} 
+ lines + contamination~model)}, 
where {\scshape borus02} represents a reprocessed component. The intrinsic cutoff power law includes line-of-sight 
absorption and losses that occurred due to Compton-scattering. 
While fitting the joint spectra, we allowed all the parameters ({\eg} column density, spectral index,
inclination angle and opening angle of the torus) to vary but tied across epochs.
Similar to the {\scshape MYTorus} model, we see residuals at soft energies ($<$ 10 keV) that can be accounted for by adding narrow Gaussians for the emission lines at various energies between 3.0 to 7.5 keV.  
\par
\begin{table*}
\centering
\begin{adjustbox}{width=\textwidth}
\renewcommand{\arraystretch}{1.25}
\begin{threeparttable}
\caption{The joint fit spectral parameters using {\scshape borus02} model}
\begin{tabular}{ccccccccccc}
\hline
Parameters & 1998-03-13 & 2001-01-07 & 2006-07-21 & 2013-01-25 & 2013-02-02 & 2013-02-03 & 2013-02-05 & 2016-08-23  & 2020-01-28 & 2020-11-26 \\
    & ($Beppo$) & ($Beppo$) & ($Suzaku$) & ($Nu$)   &   ($Nu$)   & ($X+Nu$)    & ($Nu$) & ($X+Nu$)  &
    ({${AstroSat}$}) & ($Nu$)     \\
    &    (1)     &    (2)    &    (3)    &  (4)   &  (5)   & (6 ) & (7)  & (8)   &  (9)   &  (10)   \\
\hline
\multicolumn{11}{c}{~Model: {\scshape borus02} + lines + contamination}  \\
\hline
\multicolumn{11}{c}{\underline{case~(i) : all parameters varied but tied across epochs}}  \\
$\Gamma$                   & \multicolumn{10}{c}{$2.39^{+0.04}_{-0.01}$}  \\
$N_{\rm H,LOS}$           & \multicolumn{10}{c}{$6.10^{+0.33}_{-0.04}$}  \\
$N_{\rm H,tor}$           & \multicolumn{10}{c}{$14.05^{+0.98}_{-0.82}$} \\
${\theta}_{\rm tor}$       & \multicolumn{10}{c}{$73.8^{+0.1}_{-0.1}$} \\
${\theta}_{\rm incl}$      & \multicolumn{10}{c}{$80.8^{+0.1}_{-0.1}$} \\
$f_{\rm cov}$              & \multicolumn{10}{c}{$0.28^{+0.01}_{-0.01}$} \\
$A_{\rm Z}$ = $A_{\rm S}$  & \multicolumn{10}{c}{$13.58^{+1.27}_{-1.36}$} \\
$E_{\rm cut}$              & \multicolumn{10}{c}{$108.3^{+1.2}_{-2.1}$} \\
${\chi}_{r}$(dof)          & \multicolumn{10}{c}{$1.138~(8105)$} \\
$C_{\rm XMM}$  &         &       &         &      &     &  $1.0^{f}$ &      & $0.94^{+0.01}_{-0.01}$ &     &     \\
$C_{\rm NuA}$  &         &       &   & $1.29^{+0.01}_{-0.01}$ & $1.27^{+0.01}_{-0.01}$ & $1.24^{+0.01}_{-0.01}$ & $1.20^{+0.01}_{-0.01}$ &  $1.34^{+0.01}_{-0.01}$   &     & $1.21^{+0.01}_{-0.01}$ \\
$C_{\rm NuB}$   &          &       &     &  $1.34^{+0.01}_{-0.01}$ & $1.31^{+0.01}_{-0.01}$ & $1.26^{+0.01}_{-0.01}$ & $1.23^{+0.01}_{-0.01}$ & $1.40^{+0.01}_{-0.01}$ &      & $1.22^{+0.01}_{-0.01}$    \\
$C_{\rm XIS0}$ & $0.84^{+0.01}_{-0.01}$ (ME) & $0.77^{+0.01}_{-0.01}$ (ME) & $0.97^{+0.01}_{-0.01}$ &         &     &   &      &     & $0.69^{+0.05}_{-0.05}$ (S) &     \\
$C_{\rm XIS1}$ &           &       & $0.99^{+0.01}_{-0.01}$ &         &     &   &     &     &     &     \\
$C_{\rm XIS2}$ &           &       & $0.93^{+0.01}_{-0.01}$ &         &     &   &     &     &     &     \\
$C_{\rm XIS3}$ &           &       & $0.93^{+0.01}_{-0.01}$ &         &     &   &     &     &     &     \\
$C_{\rm PIN}$  & $1.20^{+0.02}_{-0.02}$ (P) & $1.04^{+0.02}_{-0.02}$ (P) & $1.45^{+0.02}_{-0.02}$ &         &     &   &      &     &  $0.88^{+0.05}_{-0.05}$ (LX)   &     \\
$C_{\rm GSO}$  &           &       & $1.52^{+0.24}_{-0.24}$ &         &     &   &    &     & $1.55^{+0.22}_{-0.22}$ (CZ) &     \\
\hline
\multicolumn{11}{c}{\underline{case~(ii) : only normalisation is untied across epochs}}  \\
$\Gamma$                   & \multicolumn{10}{c}{$2.39^{+0.05}_{-0.02}$}  \\
$N_{\rm H,LOS}$       & \multicolumn{10}{c}{$6.16^{+0.23}_{-0.16}$} \\
$N_{\rm H,tor}$           & \multicolumn{10}{c}{$14.05^{+0.91}_{-0.77}$} \\
${\theta}_{\rm tor}$       & \multicolumn{10}{c}{$73.7^{+0.2}_{-0.1}$} \\
${\theta}_{\rm incl}$      & \multicolumn{10}{c}{$80.8^{+0.1}_{-0.1}$} \\
$f_{\rm cov}$              & \multicolumn{10}{c}{$0.28^{+0.01}_{-0.01}$} \\
$A_{\rm Z}$ = $A_{\rm S}$ & $14.09^{+2.13}_{-1.41}$ & $13.19^{+1.51}_{-1.00}$ & $14.17^{+2.36}_{-1.57}$ & $13.26^{+0.33}_{-1.40}$ & $13.21^{+1.49}_{-1.49}$ & $13.92^{+1.35}_{-1.35}$ & $13.52^{+1.47}_{-1.47}$ &  $15.02^{+0.62}_{-1.57}$ & $17.97^{+2.31}_{-2.00}$ & $13.65^{+0.30}_{-1.75}$ \\
$E_{\rm cut}$              & \multicolumn{10}{c}{$105.7^{+2.7}_{-2.2}$} \\
${\chi}_{r}$(dof)          & \multicolumn{10}{c}{$1.134~(8096)$} \\
\hline
\multicolumn{11}{c}{\underline{case~(iii) : only $N_{\rm H,LOS}$ is untied across epochs}}  \\
$\Gamma$                   & \multicolumn{10}{c}{$2.40^{+0.02}_{-0.01}$}  \\
$N_{\rm H,LOS}$ & $5.85^{+1.76}_{-0.81}$ & $5.22^{+0.90}_{-0.68}$ & $4.19^{+0.16}_{-0.14}$ & $6.24^{+0.18}_{-0.15}$ & $6.32^{+0.19}_{-0.16}$ & $6.30^{+0.26}_{-0.21}$    & $6.15^{+0.20}_{-0.16}$ & $5.84^{+0.17}_{-0.14}$ & $5.37^{+2.66}_{-0.72}$ & $6.22^{+0.13}_{-0.12}$ \\
$N_{\rm H,tor}$           & \multicolumn{10}{c}{$14.05^{+0.98}_{-1.01}$} \\
${\theta}_{\rm tor}$       & \multicolumn{10}{c}{$73.7^{+0.2}_{-0.1}$} \\
${\theta}_{\rm incl}$      & \multicolumn{10}{c}{$80.8^{+0.2}_{-0.1}$} \\
$f_{\rm cov}$              & \multicolumn{10}{c}{$0.28^{+0.01}_{-0.01}$} \\
$A_{\rm Z}$ = $A_{\rm S}$       & \multicolumn{10}{c}{$14.26^{+0.26}_{-1.01}$} \\
$E_{\rm cut}$              & \multicolumn{10}{c}{$110.5^{+2.9}_{-2.0}$} \\
${\chi}_{r}$(dof)          & \multicolumn{10}{c}{$1.117~(8096)$} \\
\hline
\multicolumn{11}{c}{\underline{case~(iv) : $N_{\rm H,LOS}$ and normalisations untied across epochs}}  \\
$\Gamma$                   & \multicolumn{10}{c}{$2.40^{+0.04}_{-0.02}$}  \\
$N_{\rm H,LOS}$ & $5.87^{+1.18}_{-1.00}$ & $5.15^{+1.00}_{-0.68}$ & $4.19^{+0.42}_{-0.30}$ & $6.24^{+0.22}_{-0.18}$ & $6.39^{+0.28}_{-0.25}$ & $6.38^{+0.32}_{-0.22}$    & $6.13^{+0.31}_{-0.21}$ & $5.89^{+0.28}_{-0.21}$ &
$5.69^{+0.40}_{-0.29}$ & $6.28^{+0.27}_{-0.20}$ \\
$N_{\rm H,tor}$           & \multicolumn{10}{c}{$14.05^{+0.86}_{-0.93}$} \\
${\theta}_{\rm tor}$       & \multicolumn{10}{c}{$73.7^{+0.1}_{-0.1}$} \\
${\theta}_{\rm incl}$      & \multicolumn{10}{c}{$80.8^{+0.1}_{-0.1}$} \\
$f_{\rm cov}$              & \multicolumn{10}{c}{$0.28^{+0.01}_{-0.02}$} \\
$A_{\rm Z}$ = $A_{\rm S}$ & $14.88^{+1.51}_{-1.44}$ & $13.59^{+1.19}_{-1.13}$ & $14.82^{+2.01}_{-1.71}$ & $14.05^{+0.71}_{-0.74}$ & $14.62^{+1.40}_{-1.58}$ & $14.81^{+1.11}_{-0.77}$ & $13.88^{+1.27}_{-1.63}$ &  $14.60^{+1.34}_{-1.62}$ & $16.93^{+3.31}_{-2.84}$ & $14.68^{+0.43}_{-0.96}$ \\
$E_{\rm cut}$              & \multicolumn{10}{c}{$109.4^{+4.8}_{-2.2}$} \\
${\chi}_{r}$(dof)          & \multicolumn{10}{c}{$1.117~(8087)$} \\
\hline
\end{tabular}
\label{tab:JointBorusFits}
\begin{tablenotes}
\normalsize
\item Notes - The abbreviation for spectral parameters and their units are same as mentioned in Table~\ref{tab:MYToFits}.
\end{tablenotes}
\end{threeparttable}
\end{adjustbox}
\end{table*}
We find that the {\scshape borus02} model provides an acceptable fit to the multi-epoch spectra
with reduced ${\chi}^{2}$ = 1.138 for 8105 dof (see case~(i) in Table~\ref{tab:JointBorusFits}).
We note that, untying normalisations across epochs renders a slight improvement in the fit statistics
with reduced ${\chi}^{2}$ = 1.134 for 8096 dof. Although, keeping variable $N_{\rm H,LOS}$ across epochs
gives further improvement in the fit statistics with reduced ${\chi}^{2}$ = 1.117 for 8096 dof
(see case~(iii) in Table~\ref{tab:JointBorusFits}). Notably, we obtain nearly the same fit statistics and
parameters if both normalisations and $N_{\rm H,LOS}$ are kept variable across epochs
(see case~(iv) in Table~\ref{tab:JointBorusFits}). Also, untying other parameters ({\eg} photon index,
${\theta}_{\rm tor}$, ${\theta}_{\rm incl}$) across epochs gives no significant improvement in the fit statistics.
Thus, the best fit of multi-epoch joint spectra
suggests for a variable line-of-sight column density.
We note that, similar to the {\scshape MYTorus} model, the {\scshape borus02} model too shows that
the multi-epoch spectra of AGN can be characterised by
a steep power law ($\Gamma~=~2.40^{+0.04}_{-0.02}$) emission piercing through Compton-thick
line-of-sight column density of $4.19^{\rm +0.42}_{-0.30}-6.39^{+0.28}_{-0.25}~\times~ 10^{24}~{\rm cm}^{-2}$
with a nearly edge-on torus (${\theta}_{\rm incl}$ $\simeq$ $80.8^{+0.1}_{-0.1}$).
However, unlike the {\scshape MYTorus} model, the {\scshape borus02} model allows us to constrain the globally
averaged column density ({$N_{\rm H,tor}$ = $14.05^{+0.86}_{-0.93}~\times~10^{24}~{\rm cm}^{-2}$}),
torus opening angle (${\theta}_{\rm tor}$) $\simeq$ {$73^{\circ}.7$} and
covering factor ($f_{\rm cov}$ = {$0.28^{+0.01}_{-0.02}$}) (see Table~\ref{tab:JointBorusFits}).
Thus, circumnuclear material in Circinus can be described as a thin torus with an opening angle of 73$^{\circ}$.7, rendering a relatively low covering factor of 0.28.
\subsection{Spectral modelling with {\scshape UXCLUMPY} model}
\label{sec:UXClumpy}
We note that both the {\scshape MYTorus} and {\scshape borus02} models consider reprocessor of uniform density,
however, circumnuclear material can be clumpy \citep{Nenkova08a}.
To examine the viability of clumpy circumnuclear material, we perform spectral fitting with
the {\scshape UXCLUMPY} model \citep{Buchner19} which considers clumpy structure of the reprocessing matter
consists of small (angle subtended at the black hole in the range $0'.1 - 1^{\circ}$) spherical clouds each 
having constant density. 
The clouds are distributed in an axis-symmetric geometry with a decreasing number towards the pole 
from the equatorial plane following a Gaussian function $N = N_0{\rm exp} \left\{ -(\beta/\sigma)^m \right\}$, 
where $N$ denotes the number of clouds seen along the line-of-sight, $N_0$ is the number of clouds at the equatorial 
plane, $\beta$ is the inclination angle towards the pole from the equatorial plane, 
$\sigma$ is the angular width of the distribution (the torus scale height) and varies in the 
range $6^{\circ} - 90^{\circ}$.   
In addition to an axisymmetric clumpy torus, 
this model considers a Compton-thick reflector near the corona, which can be interpreted as 
part of the dust-free broad-line region or a warped disk. 
The clumpy reprocessing material considered in this model results in the leakage of soft photons
along the Compton-thick sight lines.
Similar to the other models, the {\scshape UXClumpy} model considers transmitted component, 
Compton-scattered component and fluorescent emission lines in a self-consistent manner. 
Although, unlike previous models, the {\scshape UXCLUMPY} model considers line-of-sight column density that can vary
in the range of 10$^{20}$ cm$^{-2}$ to 10$^{26}$ cm$^{-2}$.
\par
We fitted the joint spectra of all ten epochs using a model defined as
\textsc{const{*tbabs*(UXCLUMPY\_cutoff + lines + contamination)}}, where the {\scshape UXCLUMPY\_cutoff}
represents {\scshape UXCLUMPY} model with a high energy cutoff in the range of 60 keV to 400 keV. 
The constant parameter and {\scshape tbabs} account for the cross-normalisation factor, and galactic column 
density, respectively.    
While performing the fit, we kept all the parameters of the {\scshape UXCLUMPY} model free but tied across epochs.
We noticed that often torus inclination angle becomes very low, resulting in an unstable fit. 
Therefore, we fixed the torus inclination angle to 80$^{\circ}$ based on the values obtained
from the {\scshape MYTorus} and {\scshape borus02} models.
\par
\begin{figure}
\centering
\includegraphics[angle=-90, width=\columnwidth, trim={0.0cm 0.0cm 0.0cm 0.0cm}, clip]{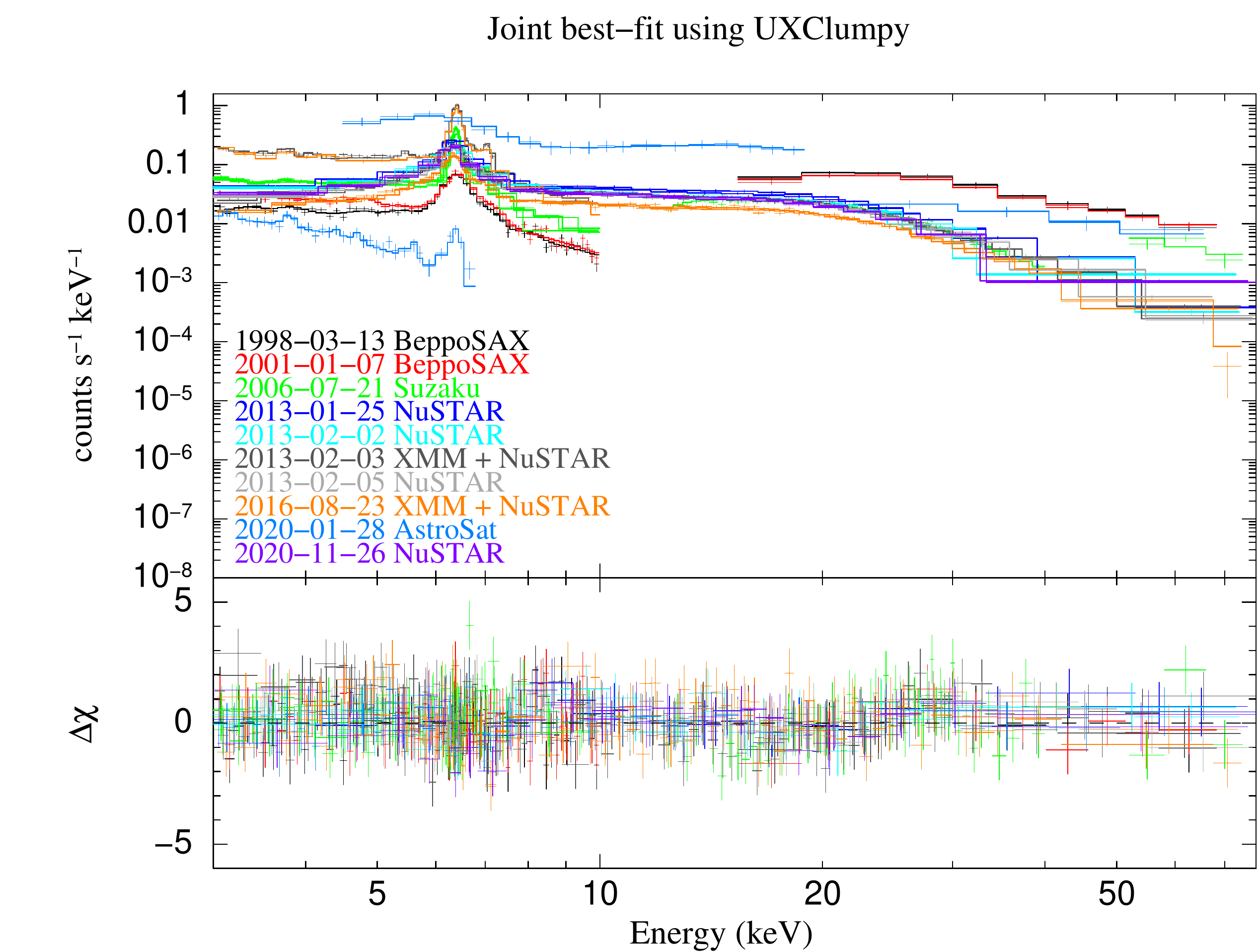}
\caption{The multi-epoch spectra jointly fitted with {\scshape UXCLUMPY} model. Residuals seen around 6$-$7 keV can be attributed to Fe K$\alpha$ and K$\beta$ emission lines detected with different instruments of differing energy resolution. For a better display of residuals all the spectra are rebinned with a minimum of 50 to 200 counts per bin.}
\label{fig:multiepochSpec}
\end{figure}

\begin{table*}
\centering
\begin{adjustbox}{width=\textwidth}
\renewcommand{\arraystretch}{1.25}
\begin{threeparttable}
\caption{The joint fit spectral parameters using UXCLUMPY model}
\begin{tabular}{ccccccccccc}
\hline
Parameters & 1998-03-13 & 2001-01-07 & 2006-07-21 & 2013-01-25 & 2013-02-02 & 2013-02-03 & 2013-02-05 & 2016-08-23  & 2020-01-28 & 2020-11-26 \\
    & ($Beppo$) & ($Beppo$) & ($Suzaku$) & ($Nu$)   &   ($Nu$)   & ($X+Nu$)  & ($Nu$) & ($X+Nu$)  &  ({${AstroSat}$}) & ($Nu$)     \\
      &    (1)     &    (2)    &    (3)    &  (4)   &  (5)   & (6 ) & (7)  & (8)   &  (9)   &  (10)   \\
\hline
\multicolumn{11}{c}{~Model: {\scshape UXCLUMPY} + lines + contamination}  \\
\hline
\multicolumn{11}{c}{\underline{case~(i) : all parameters varied but tied across epochs}}  \\
$\Gamma$                   & \multicolumn{10}{c}{$2.05^{+0.01}_{-0.02}$} \\
$N_{\rm H,LOS}$           & \multicolumn{10}{c}{$8.91^{+0.14}_{-0.08}$} \\
${\theta}_{\rm incl}$      & \multicolumn{10}{c}{$80^{f}$} \\
${\sigma}_{\rm tor}$       & \multicolumn{10}{c}{$7.0^{+0.4}_{\rm peg}$} \\
${\rm CTKCover}$           & \multicolumn{10}{c}{$0.32^{+0.01}_{-0.01}$} \\
$A_{\rm Z}$ = $A_{\rm S}$ & \multicolumn{10}{c}{$0.80^{+0.01}_{-0.03}$} \\
$E_{\rm cut}$              & \multicolumn{10}{c}{$400^{\rm peg}_{-37.4}$} \\
${\chi}_{r}$(dof)          & \multicolumn{10}{c}{$1.165~(8106)$} \\
$C_{\rm XMM}$  &         &       &         &      &     &  $1.0^{f}$ &      & $0.97^{+0.02}_{-0.01}$ &     &     \\
$C_{\rm NuA}$  &         &       &   & $1.19^{+0.01}_{-0.01}$ & $1.16^{+0.01}_{-0.02}$ & $1.16^{+0.01}_{-0.01}$ & $1.11^{+0.01}_{-0.01}$ &  $1.25^{+0.02}_{-0.02}$   &     & $1.11^{+0.01}_{-0.01}$ \\
$C_{\rm NuB}$   &          &       &     &  $1.24^{+0.01}_{-0.01}$ & $1.20^{+0.01}_{-0.03}$ & $1.19^{+0.01}_{-0.01}$ & $1.14^{+0.01}_{-0.01}$ & $1.30^{+0.02}_{-0.02}$ &      & $1.12^{+0.01}_{-0.01}$    \\
$C_{\rm XIS0}$ & $0.87^{+0.02}_{-0.02}$ (ME) & $0.80^{+0.02}_{-0.02}$ (ME) & $1.04^{+0.01}_{-0.01}$ &         &     &   &      &     & $0.73^{+0.05}_{-0.05}$ (S) &     \\
$C_{\rm XIS1}$ &           &       & $1.08^{+0.01}_{-0.01}$ &         &     &   &     &     &     &     \\
$C_{\rm XIS2}$ &           &       & $1.00^{+0.01}_{-0.01}$ &         &     &   &     &     &     &     \\
$C_{\rm XIS3}$ &           &       & $1.00^{+0.01}_{-0.01}$ &         &     &   &     &     &     &     \\
$C_{\rm PIN}$  & $1.10^{+0.02}_{-0.02}$ (P) & $0.96^{+0.02}_{-0.04}$ (P)    &  $1.34^{+0.02}_{-0.02}$ &         &     &   &      &     &  $0.84^{+0.06}_{-0.06}$ (LX)   &     \\
$C_{\rm GSO}$  &           &       & $1.48^{+0.24}_{-0.24}$ &         &     &   &    &     & $1.44^{+0.24}_{-0.24}$ (CZ) &     \\ \hline
\multicolumn{11}{c}{\underline{case~(ii) : only normalisation is untied across epochs}}  \\
$\Gamma$                   & \multicolumn{10}{c}{$2.05^{+0.01}_{-0.02}$} \\
$N_{\rm H,LOS}$       & \multicolumn{10}{c}{$8.95^{+0.19}_{-0.10}$} \\
${\theta}_{\rm incl}$      & \multicolumn{10}{c}{$80^{f}$} \\
${\sigma}_{\rm tor}$       & \multicolumn{10}{c}{$7.0^{+0.4}_{\rm peg}$} \\
${\rm CTKCover}$           & \multicolumn{10}{c}{$0.32^{+0.02}_{-0.001}$} \\
$A_{\rm Z}$ = $A_{\rm S}$ & $0.77^{+0.01}_{-0.03}$ & $0.63^{+0.01}_{-0.02}$ & $0.83^{+0.01}_{-0.03}$ & $0.79^{+0.03}_{-0.02}$ & $0.79^{+0.01}_{-0.02}$ & $0.79^{+0.02}_{-0.03}$ & $0.80^{+0.02}_{-0.02}$ & $0.82^{+0.02}_{-0.03}$ & $0.91^{+0.06}_{-0.06}$ & $0.80^{+0.02}_{-0.02}$  \\
$E_{\rm cut}$              & \multicolumn{10}{c}{$400^{\rm peg}_{-22.0}$} \\
${\chi}_{r}$(dof)          & \multicolumn{10}{c}{$1.164~(8097)$} \\
\hline
\multicolumn{11}{c}{\underline{case~(iii) : only $N_{\rm H,LOS}$ is untied across epochs}}  \\
$\Gamma$                   & \multicolumn{10}{c}{$2.06^{+0.01}_{-0.01}$} \\
$N_{\rm H,LOS}$  & $7.84^{+0.24}_{-0.21}$ & $12.01^{+0.30}_{-0.26}$ & $6.97^{+0.07}_{-0.05}$ & $8.84^{+0.19}_{-0.11}$ & $8.95^{+0.43}_{-0.29}$ & $8.99^{+0.30}_{-0.25}$ & $8.59^{+0.29}_{-0.28}$ & $6.54^{+0.14}_{-0.13}$ & $4.46^{+1.42}_{-0.94}$ & $8.84^{+0.09}_{-0.08}$ \\
${\theta}_{\rm incl}$      & \multicolumn{10}{c}{$80^{f}$} \\
${\sigma}_{\rm tor}$       & \multicolumn{10}{c}{$7.0^{+0.2}_{\rm peg}$} \\
${\rm CTKCover}$           & \multicolumn{10}{c}{$0.31^{+0.005}_{-0.001}$} \\
$A_{\rm Z}$ = $A_{\rm S}$       & \multicolumn{10}{c}{$0.80^{+0.01}_{-0.002}$} \\
$E_{\rm cut}$              & \multicolumn{10}{c}{$400^{\rm peg}_{-14.3}$} \\
${\chi}_{r}$(dof)          & \multicolumn{10}{c}{$1.159~(8097)$} \\
\hline
\multicolumn{11}{c}{\underline{case~(iv) : $N_{\rm H,LOS}$ and normalisations untied across epochs}}  \\
$\Gamma$                   & \multicolumn{10}{c}{$2.08^{+0.01}_{-0.01}$} \\
$N_{\rm H,LOS}$  & $6.42^{+0.38}_{-0.36}$ & $4.70^{+0.46}_{-0.44}$ & $4.32^{+0.04}_{-0.03}$ & $8.63^{+0.18}_{-0.14}$ & $8.93^{+0.24}_{-0.31}$ & $9.26^{+0.75}_{-0.53}$ & $8.36^{+0.81}_{-0.67}$ & $5.38^{+0.21}_{-0.21}$ & $4.13^{+2.23}_{-1.07}$ & $8.84^{+0.16}_{-0.15}$ \\
${\theta}_{\rm incl}$      & \multicolumn{10}{c}{$80^{f}$} \\
${\sigma}_{\rm tor}$       & \multicolumn{10}{c}{$7.1^{+0.6}_{\rm peg}$} \\
${\rm CTKCover}$           & \multicolumn{10}{c}{$0.33^{+0.01}_{-0.01}$} \\
$A_{\rm Z}$ = $A_{\rm S}$ & $0.77^{+0.02}_{-0.02}$ & $0.61^{+0.02}_{-0.02}$ & $0.82^{+0.04}_{-0.01}$ & $0.87^{+0.02}_{-0.01}$ & $0.88^{+0.02}_{-0.01}$ & $0.86^{+0.02}_{-0.01}$ & $0.87^{+0.02}_{-0.01}$ & $0.69^{+0.02}_{-0.01}$ & $0.70^{+0.27}_{-0.07}$ & $0.88^{+0.01}_{-0.01}$  \\
$E_{\rm cut}$              & \multicolumn{10}{c}{$400^{\rm peg}_{-17.8}$} \\
${\chi}_{r}$(dof)          & \multicolumn{10}{c}{$1.152~(8088)$} \\
\hline
\end{tabular}
\label{tab:JointUXClumpyFits}
\begin{tablenotes}
\normalsize
\item Notes - The abbreviation for spectral parameters and their units are the same as mentioned in Table~\ref{tab:MYToFits}.
\end{tablenotes}
\end{threeparttable}
\end{adjustbox}
\end{table*}
Similar to our previous spectral modellings, we attempted to fit the multi-epoch spectra with
the {\scshape UXCLUMPY} model using different scenarios.
The spectral fitted parameters based on the {\scshape UXCLUMPY} model are listed in Table~\ref{tab:JointUXClumpyFits}.
We plot the best$-$fitted joint spectra and residuals in Figure~\ref{fig:multiepochSpec}.
We find that the best fit is achieved when both $N_{\rm H,LOS}$ and normalisations
are varied across epochs (see case (iv) in Table~\ref{tab:JointUXClumpyFits}).
We also tried varying other parameters ({\eg} photon index, ${\sigma}_{\rm tor}$, and {\tt CTKCover}) but
found no further significant improvement in the fit statistics.
We note that similar to the {\scshape MYTorus} and {\scshape borus02} models, the spectral modelling with
the {\scshape UXCLUMPY} too reveals
a steep spectrum ($\Gamma~=~2.08^{+0.01}_{-0.01}$) AGN obscured with Compton-thick column densities.
The line-of-sight column density ($N_{\rm H,LOS}$) changes from {$4.13^{+2.23}_{-1.07}~\times~10^{24}~{\rm cm}^{-2}$} to $9.26^{+0.75}_{-0.53}~\times~10^{24}~{\rm cm}^{-2}$ across epochs.
A significant change in $N_{\rm H,LOS}$ from one to other epochs
at various timescales is consistent with a clumpy obscuring
material wherein the obscuring clouds can move in or out of the line-of-sight.
A detailed discussion on the implications of variable $N_{\rm H,LOS}$ is given in Section~\ref{sec:Discussion}.
\par
We note that the angular dispersion of clouds in the torus (${\sigma}_{\rm tor}$) is nearly
7$^{\circ}$ (with lower limit pegging to the hard limit of 6$^{\circ}$), and hence,
${\sigma}_{\rm tor}$ parameter indicates a thin torus.
However, our spectral modelling always requires an inner ring of Compton-thick material to fully account for 
the scattered components. The covering factor of the inner Compton-thick reflector 
({\tt CTKCover}) is {$0.33^{+0.01}_{-0.01}$}.
Both {\tt CTKCover} and ${\sigma}_{\rm tor}$ parameters provide constraints on the geometry of the circumnuclear
material. The implications of these parameters are discussed in Section~\ref{sec:Discussion}. 
We point out that our spectral parameters from the {\scshape UXCLUMPY} model are 
broadly consistent with those obtained 
by \cite{Buchner19}, although we find a somewhat steeper photon index and require no warm mirror 
component, which is a scattered component arriving to the observer without encountering heavy absorption 
from the torus. One of the main differences in our spectral fitting is the inclusion of the contamination 
model. 
\section {Discussion}
\label{sec:Discussion}
\subsection{Constraints on the geometry and covering factor}
With the modelling of multi-epoch spectra of Circinus, we find the evidence for reprocessing
material distributed in the form of torus around AGN. The best-fitted models favour a nearly edge-on torus with an
inclination angle of 77$^{\circ}$ $-$ 81$^{\circ}$, which is consistent with
the findings at other wavelengths. For instance, IR interferometric observations revealed a
circumnuclear dusty disc component with inclination angle of $>$ 75$^{\circ}$ \citep[see][]{Tristram14,Isbell22}.
In sub-millimeter wavelengths, both continuum as well as CO(3-2), [C {\scriptsize I}](1-0) emission line maps obtained with
the Atacama Large Millimeter/Submillimeter Array (ALMA) showed a circumnuclear disk having the size of a few tens
of parsec and inclination angle of $>$ 70$^{\circ}$. Also, H$_2$O maser disc is found to be edge-on with an
inclination angle close to 90$^{\circ}$ \citep{Greenhill03}. Therefore, a multi-phase circumnuclear
material is plausibly distributed in the form of a torus that transforms into a disc at inner regions around AGN.
\par
\begin{figure}
\centering
\includegraphics[angle=-90, width=\columnwidth, trim={0.0cm 0.0cm 0.0cm 0.0cm}, clip]{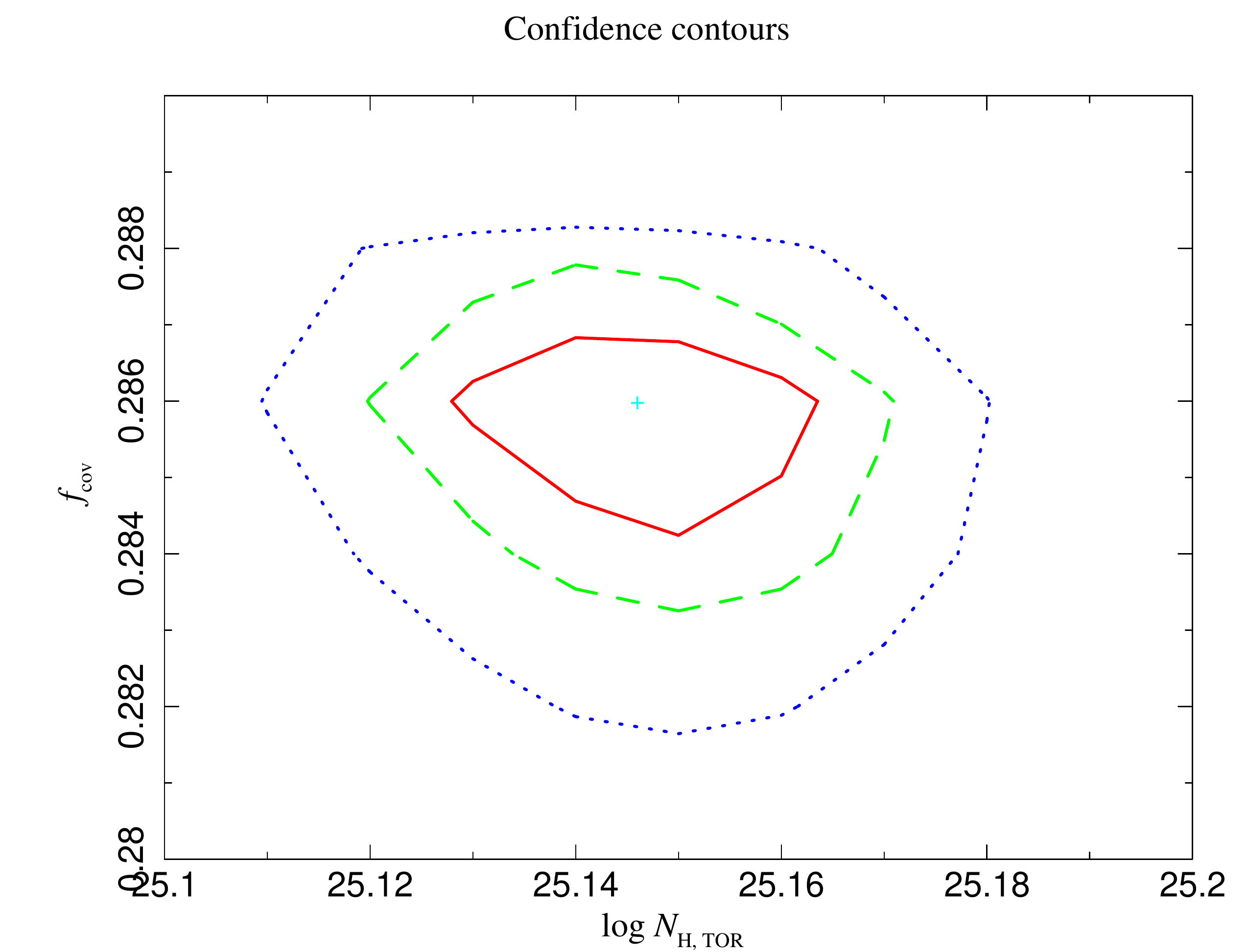}
\caption{The confidence contours based on the joint spectral fitting with the {\scshape borus02} model.
The confidence contours from inward to outward represent 1$\sigma$, 2$\sigma$ and 3$\sigma$ significant level.}
\label{fig:CovVsNh}
\end{figure}

Further, we attempted to place constraints on the covering factor of circumnuclear reprocessing material.
The {\scshape MYTorus} model yielding a good fit for the multi-epoch 
spectra assumes a fixed value of covering factor ($f_{\rm cov}$) of 0.5. However,
covering factor should not be fixed to a particular value in a prior.
Therefore, we applied the {\scshape borus02} model, which considers $f_{\rm cov}$ as a free parameter.
We showed that our multi-epoch spectra could be jointly fitted with the {\scshape borus02} model resulting covering factor of 
nearly 0.28 (see Table~\ref{tab:JointBorusFits}). 
The confidence contours between the covering factor ($f_{\rm cov}$) and average column density 
($N_{\rm H,tor}$) show that both the parameters derived from the multi-epoch joint fit are well 
constrained (see Figure~\ref{fig:CovVsNh}). 
Also, since keeping the covering factor a variable parameter across epochs does not improve the spectral fit,
and hence, we conclude that the overall geometry of the torus remains nearly the same across all epochs.
\par 
We recall that the {\scshape borus02} model allows us to estimate the covering factor, but it 
considers a uniform reprocessing material. Keeping this limitation in mind, we used the {\scshape UXCLUMPY} model, which considers 
clumpy reprocessing material with clouds having a Gaussian distribution along the direction
perpendicular to the torus equatorial plane. This model also considers an inner ring of Compton-thick
material to explain Compton hump, if needed. We find that our spectra are well fitted
with the {\scshape UXCLUMPY} model, which gives angular dispersion of nearly 7$^{\circ}$.
Thus, a low value of angular dispersion infers a thin torus. 
We note that a direct comparison of covering factors obtained in the {\scshape borus02}
and the {\scshape UXCLUMPY} models is not possible as they assume different geometries for the reprocessing material. Although, both models infer that the torus is likely to be thin with a relatively low covering factor.
Further, we point out that an inner ring of Compton-thick reflector is needed in the {\scshape UXCLUMPY} model
to well describe the spectra of all epochs. The covering factor of the inner ring ({\tt CTKCover}) is found to be 0.33.
The inner ring may be interpreted as the wind launch site \citep{Krolik88}, a puffed-up inner rim of accretion
disk or a warped disk which tends to hide AGN when the torus or accretion disk is viewed edge-on,
as in the case of Circinus. In a recent work, \cite{Andonie22} suggested that the inner ring can plausibly depict the accretion disc and the BLR region.
\par
In the literature, there have been several attempts to probe the circumnuclear material in Circinus.
For instance, recently \cite{Ursini23} performed Imaging X-ray Polarimetry Explorer (IXPE) observations and reported
a significantly polarised (28$\pm$7 per cent degree of polarization and
a polarization angle of 18$^{\circ}$$\pm$5$^{\circ}$) $2.0-6.0$~keV soft X-ray AGN emission mostly due to
a neutral reflector. The observed polarization properties were explained by considering the reprocessing of X-ray
emission from an edge$-$on
uniform-density torus having the ratio of inner to outer radius 0.1$-$0.5 and a half-opening angle of 45$^{\circ}$~$-$~55$^{\circ}$, which infers a covering factor of nearly 0.5, a value consistent with the MYTorus model.
However, considering IR Interferometric observations of Circinus providing a direct evidence
of clumpy circumnuclear environment \citep[see][]{Tristram14,Stalevski19,Isbell22}, we
favour the geometrical parameters inferred from the models assuming clumpy circumnuclear material.
We point out that, \cite{Uematsu21} modelled $3-100$~keV broad-band X-ray spectrum of Circinus with
a clumpy torus model and suggested a Compton-thick but geometrically thin torus
with an angular width ({$\sigma$}) = {$10.3^{+0.7}_{-0.3}$} degrees,
which is consistent with our findings.
Further, covering factor can be related to the accretion rate or AGN intrinsic luminosity with
a trend of decreasing covering factor with the increase in accretion rate and AGN luminosity \citep[see][]{Buchner17,Ricci17b}. The low covering factor ($f_{\rm c}$ = 0.28) found in the Circinus can be understood
due to its high accretion rate, {\em i.e.,} $L$/$L_{\rm Edd}$ = 0.2.
Circinus hosts a luminous AGN. The absorption-corrected intrinsic {$2-10$} keV AGN luminosity
in different epochs is found to be in the range of 2.0 $-$ 3.7 $\times$ 10$^{42}$ erg s$^{-1}$,
which is also consistent with the value inferred from the infrared-X-ray correlation \citep[see,][]{Gandhi09,Asmus15}.
\begin{figure}
\centering
\includegraphics[angle=0, width=\columnwidth, trim={0.0cm 0.0cm 0.0cm 0.0cm}, clip]{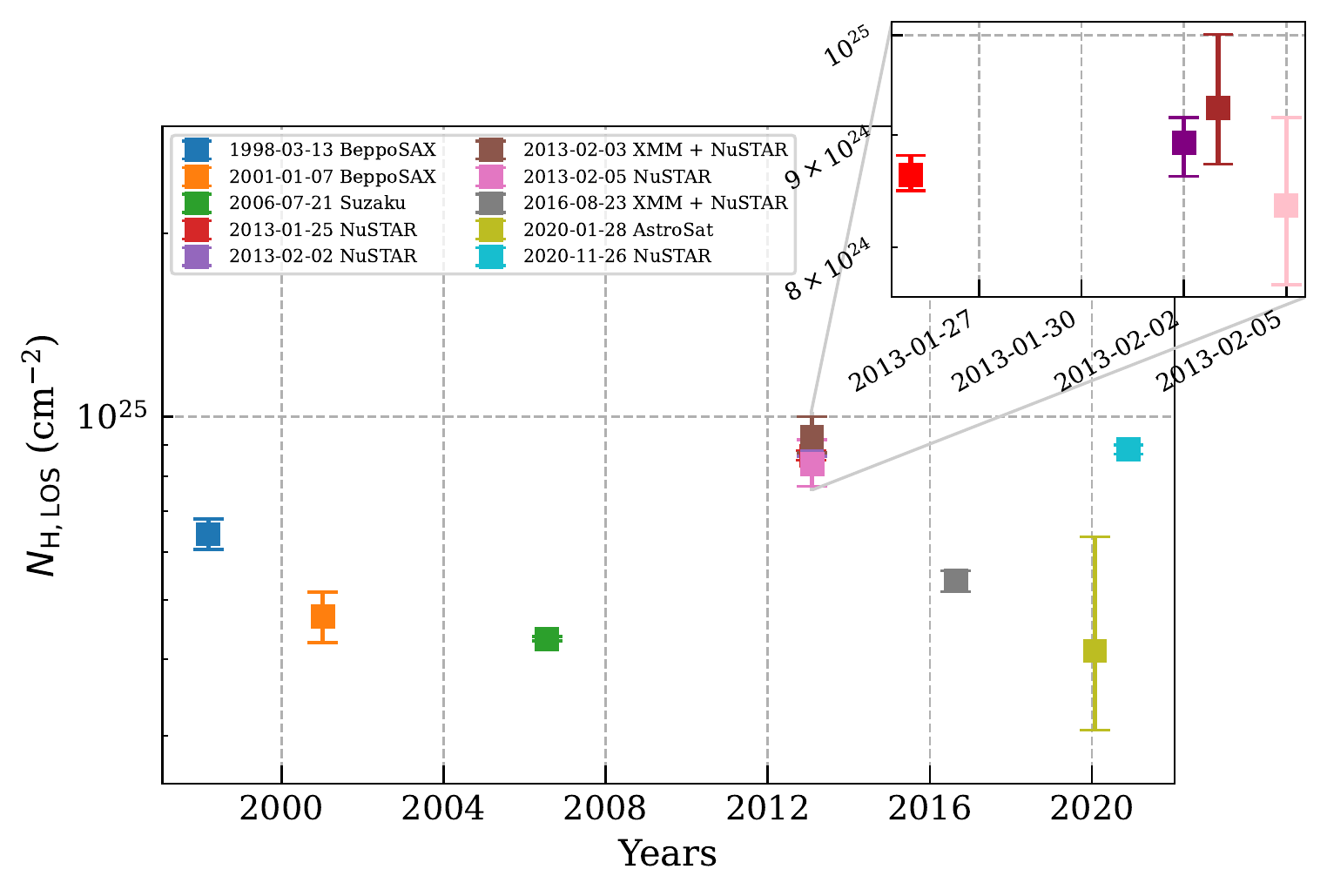}
\caption{The plot showing line-of-sight column density ($N_{\rm H,LOS}$) variability across different epochs. 
The estimates of  $N_{\rm H,LOS}$ are based on the best fit (case-iv) of joint spectral modelling
using the {\scshape UXClumpy} model.}
\label{fig:multiNH}
\end{figure}
\subsection{Variability in line-of-sight column density}

We investigated the variability
in the line-of-sight column density, which can provide a clear evidence for the clumpy circumnuclear material.
From our spectral modelling, it is evident that the measured $N_{\rm H,LOS}$ is model dependent.
Therefore, for our analysis, we prefer to use $N_{\rm H,LOS}$ measured from the {\scshape UXCLUMPY}
model owing to the fact that both {\scshape MYTorus} and {\scshape borus02} models assume a uniform-density torus.
In Figure~\ref{fig:multiNH}, we plot the line-of-sight column density ($N_{\rm H,LOS}$) for
all ten epochs using the best-fitted values obtained from the {\scshape UXCLUMPY} model.
We find that, {$N_{\rm H,LOS}$} exhibits a significant variability on years timescales.
For instance, {$N_{\rm H,LOS}$} changes from
{$6.42^{+0.38}_{-0.36}~\times~10^{24}~{\rm cm}^{-2}$}
{$4.70^{+0.46}_{-0.44}~\times~10^{24}~{\rm cm}^{-2}$} between 1998 March 13 to 2001 January 07,
{\ie} a nearly 27 per cent decrease in 3 years timescale. From 2006 July 21 to 2013 January 25, {$N_{\rm H,LOS}$} becomes nearly double, which
is 100 per cent increase in 6.5 years. The {$N_{\rm H,LOS}$} variation on months timescales is also evident.
We find that
{$N_{\rm H,LOS}$} changes from {$4.13^{+2.23}_{-1.07}~\times~10^{24}~{\rm cm}^{-2}$}
to {$8.84^{+0.16}_{-0.15}~\times~10^{24}~{\rm cm}^{-2}$} between 2020 January 28 and 2020 November 26.
The {\em NuSTAR} observations separated by one day to couple of weeks in 2013 allow us to probe
{$N_{\rm H,LOS}$} variability on shorter timescales.
One of the noticeable variations can be seen between 2013 February 03 and
2013 February 05 where {$N_{\rm H,LOS}$} changes from {$9.26^{+0.75}_{-0.53}~\times~10^{24}~{\rm cm}^{-2}$}
to {$8.36^{+0.81}_{-0.67}~\times~10^{24}~{\rm cm}^{-2}$} in two days timescales.
However, uncertainties on {$N_{\rm H,LOS}$} makes this variation only tentative.
On the shortest timescales of one day from 2013 February 02 to 2013 February 03, we find
only a marginal change ({${\Delta}N_{\rm H,LOS}$} = 0.33 {$\times$~$10^{24}~{\rm cm}^{-2}$})
in the line-of-sight column density. The significance level of change in {$N_{\rm H,LOS}$} is
low due to relatively large uncertainties. Although, we find a clear evidence of change in
{$N_{\rm H,LOS}$} on months to year timescales.
We note the our results on {$N_{\rm H,LOS}$} variation should be treated with caution as
the influence of the variability in off-nuclear contaminating sources, albeit accounted by using contamination model, cannot be completely ruled out.
We also point out that several studies in the literature have demonstrated the change in the line-of-sight column density on timescales ranging from hours to months as clouds of the clumpy material pass in and
out of the observer's line of sight \citep[see][]{Risaliti10,Ricci16}.
Therefore, similar to other nearby CT-AGN ({\eg} MRK 3; \citealp{Guainazzi16}, NGC 1068; \citealp{Zaino20}),
our study demonstrates the plausible
variability in $N_{\rm H,LOS}$ in Circinus on various timescales ranging from days to months to years.
\subsection{Location of obscuring clouds}
To know the location of obscuring cloud, we follow the method proposed by 
\cite{Risaliti02,Risaliti05} according to which the distance between the obscuring clouds and the SMBH can be estimated
using the equation given below. 
\begin{equation*}
\label{eq1}
D_{\rm cl-BH} = 600~{t^{2}_{\rm 100}}{n^{\rm 2}_{\rm 10}}{N^{-2}_{\rm H,24}}{R_{\rm S}} 
\end{equation*}
where $D_{\rm cl-BH}$ is the distance between the obscuring cloud and the black hole in units of parsec,
$t_{\rm 100}$ is time in units of 100 ks over which $N_{\rm H,LOS}$ variability observed,
$n_{\rm 10}$ is the cloud density in units
of 10$^{10}$~cm$^{-3}$ and $N_{\rm H,24}$ is the change in the line-of-sight column density in units of 10$^{24}$ cm$^{-2}$, and $R_{\rm S}$ is Schwarzschild radius.
We caution that the aforementioned equation assumes that the change in $N_{\rm H,LOS}$ is caused
due to a single cloud crossing the line-of-sight. It also assumes that the cloud is
located sufficiently close to the AGN resulting into a large coverage of the X-ray emitting AGN.
Thus, $N_{\rm H,LOS}$ variations are expected to occur only on timescales of a few days or even shorter.
Hence, to estimate the location of obscuring cloud, we use $N_{\rm H,LOS}$ variations observed only on the
short timescales of a few  days to couple of weeks.
For the sake of consistency with the assumption of the adopted method \citep[see][]{Risaliti02,Risaliti05}, we used $N_{\rm H,LOS}$ 
derived from the {\scshape UXCLUMPY} model. We estimate cloud density as $n$ = $|N_{\rm H, obs1} - N_{\rm H, obs2}|/d_{\rm corona}$
with an assumption that the change in $N_{\rm H,LOS}$ is due to a single cloud, where $N_{\rm H, obs1}$ and $N_{\rm H, obs2}$ 
are the line-of-sight column densities at two different epochs, and $d_{\rm corona}$ is the size of the corona. 
%
%
%
\begin{table*}
\centering
\begin{adjustbox}{width=\textwidth}
\begin{threeparttable}
\caption{The estimate of distance between obscuring cloud and SMBH}
\begin{tabular}{ccccccccccc}
\hline
Observation range & $t_{\rm 100~ks}$ & ${\Delta}N_{\rm H,LOS,24}$ & $n_{\rm 10,~3Rs}$ & $n_{\rm 10,~15Rs}$ &
$D_{\rm cl-BH,~3Rs}$ & $D_{\rm cl-BH,~15Rs}$ \\
&  (100~ks)   &  ($10^{24}$~cm$^{-2}$) & ($10^{10}$~cm$^{-3}$) & ($10^{10}$~cm$^{-3}$) & (pc)  & (pc)  \\ \hline
2013-01-25 to 2013-02-02 & 6.81   & 0.30 & 19.87 & 3.97 & 19.93 & 0.79 \\
2013-02-02 to 2013-02-03 & 0.94   & 0.33 & 21.85 & 4.36 & 0.38  & 0.015 \\
2013-02-03 to 2013-02-05 & 1.80   & 0.90 & 59.60 & 11.90 & 1.40  & 0.056 \\
\hline
\end{tabular}
\label{tab:Dist}
\begin{tablenotes}
\normalsize
\item Notes - $D_{\rm cl-BH,~3Rs}$  and $D_{\rm cl-BH,~15Rs}$ are the estimates of distance between obscuring cloud and SMBH using coronal size set equal to 3$R_{\rm S}$ and 15$R_{\rm S}$, respectively.
\end{tablenotes}
\end{threeparttable}
\end{adjustbox}
\end{table*}
In general, X-ray corona size ($d_{\rm corona}$) is found to be in the range of 3$R_{\rm S}$ $-$ 15$R_{\rm S}$ 
\citep[see][]{McHardy05,Fabian15,Kamraj18}. Considering $M_{\rm BH}$ = 1.7 $\times$ 10$^{6}$~$M_\odot$ \citep{Koss17} 
for Circinus, we obtained $R_{\rm S}$ (2$GM_{\rm BH}/c^{2}$) = $5.04~\times~10^{11}~{\rm cm}$.
\par
Using variability in $N_{\rm H,LOS}$ measured across different timescales, we list the estimated distance between the
obscuring clouds and the SMBH in Table~\ref{tab:Dist} for two extreme values (3$R_{\rm S}$ and 15$R_{\rm S}$) of corona size.
As expected, we find that the change in $N_{\rm H,LOS}$ on the shortest timescales is associated with the obscuring clouds located 
closest to the AGN. For instance, $N_{\rm H,LOS}$ variability on a one-day timescale between the {\em NuSTAR} observations 
taken on 2013 February 02 and 2013 February 03, provide $0.015-0.38$ parsec distance range for obscuring cloud.
Hence, the material responsible for the $N_{\rm H,LOS}$ variability on a one-day timescale is likely to be associated with the inner part of the torus. It is worth mentioning that, IR interferometric observations of Circinus have revealed dust emission
from putative torus with a projected size of 0.2$\times$1.1 pc disk-like geometry \citep[see][]{Tristram14}. 
Therefore, our constraints on the location of obscuring material based on the 
hard X-ray observations are consistent with the IR interferometric observations. 
Also, our estimate on the location of clouds in Circinus is similar to those reported
in other nearby AGN, {\eg} NGC~1068 \citep{Zaino20} and NGC~1358 \citep{Marchesi22}.   
The $N_{\rm H,LOS}$ variability on a one-week timescale seen between 2013 January 25 and 2013 February 02 suggests 
the location of clouds at a distance of 0.79 to 19.93 parsec, which may be associated with the outer part of the obscuring torus.
Also, we caution that the location of the obscuring material inferred in our analysis is only a
characteristic due to the significant errors associated with $N_{\rm H,LOS}$.   
\section{Summary}
\label{sec:Summary}
In this study, we present multi-epoch broadband X-ray spectral modelling of a nearby Compton-thick AGN in Circinus 
using all the available hard X-ray observations taken during ten different epochs in 22 years from 1998 to 2020. 
Six out of ten epochs of observations are from the {\em NuSTAR}, while the remaining four epochs of 
observations are from {\em BeppoSAX} (for two epochs), {\em Suzaku} and {\em AstroSat}.
The {\em Chandra} and {\em XMM-Newton} imaging observations of higher spatial resolution reveal the 
existence of off-nuclear sources that contaminate hard X-ray spectra. 
Therefore, prior to the broadband spectral modelling, we account for the spectral shape 
and contribution of contaminating sources. 
\par 
With an aim to constrain the geometry and structure of reprocessing material around AGN and its evolution, we 
performed broadband X-ray spectral modelling of all ten epochs using physically-motivated models,  
{\ie} {\scshape MYTorus}, {\scshape borus02} and {\scshape UXCLUMPY}. 
We find that AGN is heavily obscured by Compton-thick column densities during all the epochs.
{\scshape MYTorus} model reveals a purely reflection-dominated spectrum with a vanishing 
transmitted component. The obscuring torus is found to be nearly edge-on with an inclination angle of 77$^{\circ}$.
The {\scshape borus02} model infers a thin torus with an opening angle of nearly 73$^{\circ}$.7,
an inclination angle of 80$^{\circ}$.8 and a low covering factor of 0.28.
Interestingly, the {\scshape UXCLUMPY} model also suggests a thin torus with angular dispersion of obscuring cloud only 7$^{\circ}$ and an inner ring of Compton-thick material having
a covering factor of 0.33. We find that varying geometrical parameters (${\sigma}_{\rm tor}$, {\tt CTKCover})  across epochs renders no significant improvement in the fit statistics.
Therefore, it does seem that the overall structure at parsec scale is likely to remain unchanged.
One of the important results of our study is the tentative evidence of variable line-of-sight column density at all
timescales ranging from one day to one week to a few years. The variable line-of-sight column density supports the presence of eclipsing clouds associated with the reprocessing material at sub-parsec scales.
\section*{Acknowledgments}
%
We thank the anonymous reviewer for useful comments that helped us to improve the manuscript.
AK, VS, SV and NPSM acknowledge the support from the Department of Space Government of India. 
AK would like to thank Matteo Guainazzi for his help in resolving an issue with the {\em BeppoSAX} data products. 
CR acknowledges support from the Fondecyt Regular grant 1230345 and ANID BASAL project FB210003.
This publication uses the data from the {\em AstroSat} mission of the Indian Space Research  Organization  (ISRO),  archived  at  the  Indian  Space  Science  Data Centre (ISSDC).
This research made use of data from the {\em NuSTAR} mission, a project led by the California Institute of Technology, managed by the Jet Propulsion Laboratory, and funded by NASA, and of the NuSTARDAS jointly developed by the ASI Science Data Center (ASDC, Italy) and the California Institute of Technology (Caltech, USA). This research has made use of data obtained from the Chandra Data Archive and the Chandra Source Catalog, and software provided by the Chandra X-ray Center (CXC) in the application packages CIAO.
This research has made use of data obtained through the High Energy Astrophysics Science Archive Research Center Online Service, provided by the NASA/Goddard Space Flight Center.

\section*{Facilities}

{\em XMM-Newton}, {\em Chandra}, {\em NuSTAR}, {\em AstroSat}, {\em Suzaku} and {\em BeppoSAX}. 

\section*{Data Availability}

The {\em AstroSat} data presented in this paper are publicly available at the 
archives maintained by the ISSDC, Bengaluru: https://astrobrowse.issdc.gov.in/astro_archive/archive/Home.jsp. The {\em XMM-Newton}, {\em Chandra}, {\em NuSTAR}, {\em Suzaku} and {\em BeppoSAX} data are publicly available from 
the archives of HEASARC maintained by NASA. 
 


\bibliographystyle{mnras}
\bibliography{CircinusRef} 




\appendix
%

%
%


\label{lastpage}
\end{document}